\documentclass[aps,pre,twocolumn]{revtex4}
\usepackage{graphicx}
\usepackage{amsfonts}
\usepackage{amssymb}
\usepackage{amsmath}

\begin{document}

\title{Symmetry Breaking in Linearly Coupled Dynamical Lattices}
\author{G. Herring}
\affiliation{Department of Mathematics and Statistics, University of Massachusetts,
Amherst MA 01003-4515, USA}
\author{P.G.\ Kevrekidis}
\affiliation{Department of Mathematics and Statistics, University of Massachusetts,
Amherst MA 01003-4515, USA}
\author{B.A.\ Malomed}
\affiliation{Department of Interdisciplinary Studies, Faculty of Engineering, Tel Aviv
University, Tel Aviv 69978, Israel}
\author{R.\ Carretero-Gonz\'alez}
\affiliation{Nonlinear Dynamical Systems Group, Department of Mathematics and Statistics,
and Computational Science Research Center, San Diego State University, San
Diego CA, 92182-7720, USA}
\author{D.J.\ Frantzeskakis}
\affiliation{Department of Physics, University of Athens, Panepistimiopolis, Zografos,
Athens 15784, Greece }

\begin{abstract}
We examine one- and two-dimensional (1D and 2D) models of linearly
coupled lattices of the discrete-nonlinear-Schr{\"{o}}dinger type.
Analyzing ground states of the systems with equal powers in the
two components, we find a symmetry-breaking phenomenon beyond a
critical value of the squared $l^2$-norm. Asymmetric
states, with unequal powers in their components, emerge through a
subcritical pitchfork bifurcation, which, for  very
weakly coupled lattices, changes into a supercritical one. We identify
the stability of various solution branches. Dynamical
manifestations of the symmetry breaking are studied by simulating
the evolution of the unstable branches. The results present the
first example of spontaneous symmetry breaking in 2D
lattice solitons. This feature has no counterpart in the continuum limit,
because of the collapse instability in the latter case.
\end{abstract}

\date{Submitted to {\it Phys.~Rev.E}, April, 2007.}

\maketitle





\section{Introduction}

Dynamical lattices and their applications have become an area of
increasing interest over the past decade, as shown by a multitude
of recent reviews of the topic \cite{reviews}-\cite{sievers}. This
growth was driven by a wide array of physical realizations, in
fields as diverse as light propagation in optical waveguide arrays
\cite{optics}, dynamics of Bose-Einstein condensates (BECs)
in periodic potentials (optical lattices) \cite{bec_reviews},
micro-mechanical cantilever arrays \cite{sievers}, models of DNA
\cite{peyrard}, and others. A key model that has
been widely used and analyzed
in each of the above areas is the discrete nonlinear Schr{\"{o}}dinger (DNLS) equation
\cite{dnls}. In these applications, it emerges
either as a tight-binding approximation (as in the case of BECs trapped in optical lattices),
or at the level of an envelope-wave expansion of the underlying physical field (such as
the electromagnetic field of light in the optical systems).

One aspect of this class of discrete dynamical models which remains
perhaps less explored concerns their multi-component generalizations, which
are relevant to many fields where dynamical lattices are natural models. For
instance, in the case of waveguide arrays, one may consider settings with two
orthogonal polarizations of light \cite{meier}, or two different
wavelengths, see e.g., Ref.~\cite{hudock}. Similarly, in the BEC context, one may
consider multi-species condensates in the form of mixtures of different
hyperfine states in $^{87}$Rb \cite{myatt,dsh} and $^{23}$Na \cite{stamper},
or mixtures of different atomic species, such as Na--Rb, K--Rb, Cs--Rb, Li--Rb,
as well as Li--Cs (see, e.g., Ref.~\cite{susanto} and references therein).

While the above settings are typically modeled by systems of DNLS equations
which are coupled by nonlinear terms, such as the ones accounting for the
cross-phase modulation in optics, or collisions between atoms belonging to
different BEC species, it is also relevant to consider \emph{linearly coupled} DNLS equations
(or, in other discrete settings, linearly coupled Ablowitz-Ladik equations \cite{borisyang}).
In the optics context, such systems of linearly coupled DNLS equations are
relevant to various applications:
for example,
linear coupling occurs among the two modes inside each waveguide,
which may be induced by a twist of the waveguide
(for linear polarizations),
or by birefringence
(for circular polarizations),
or in a dual-core structure of the waveguide \cite{hudock}. On the other hand, in BECs, a
linear coupling may be imposed by an external microwave or radio-frequency
field, which can drive Rabi \cite{Rabi} or Josephson \cite{Josephson}
oscillations between populations of two different states.

\begin{figure}[t]
\includegraphics[width=6.5cm,height=5cm,angle=0,clip]{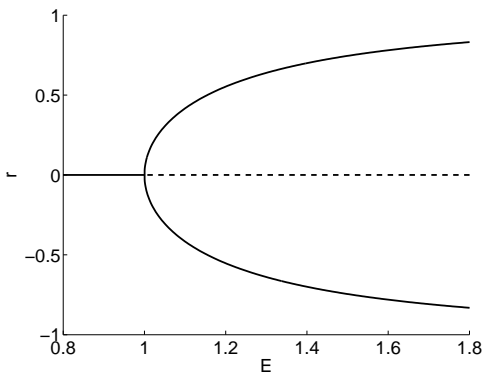}
\caption{Bifurcation diagram of solutions for the anti-continuum limit, $%
\protect\epsilon =0$;
$r$ and $E$ are the asymmetry parameter and the half of the total squared
norm, respectively
(see the text for the mathematical definitions).
The solid and dashed lines show stable and unstable solutions, respectively.}
\label{Eps0}
\end{figure}

\begin{figure}[ht]
\includegraphics[width=6.5cm,height=5cm,angle=0,clip]{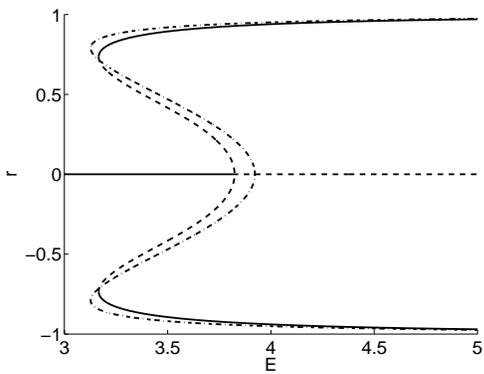}
\caption{Bifurcation diagrams for $\protect\epsilon =1.6$ in the 1D model.
The dashed-dotted line indicates solutions found by means of the variational
approximation, while solid and dashed lines show, respectively, numerically
found stable and unstable steady-state solutions.}
\label{Branches}
\end{figure}

\begin{figure}[ht]
\includegraphics[width=6.5cm,height=5cm,angle=0,clip]{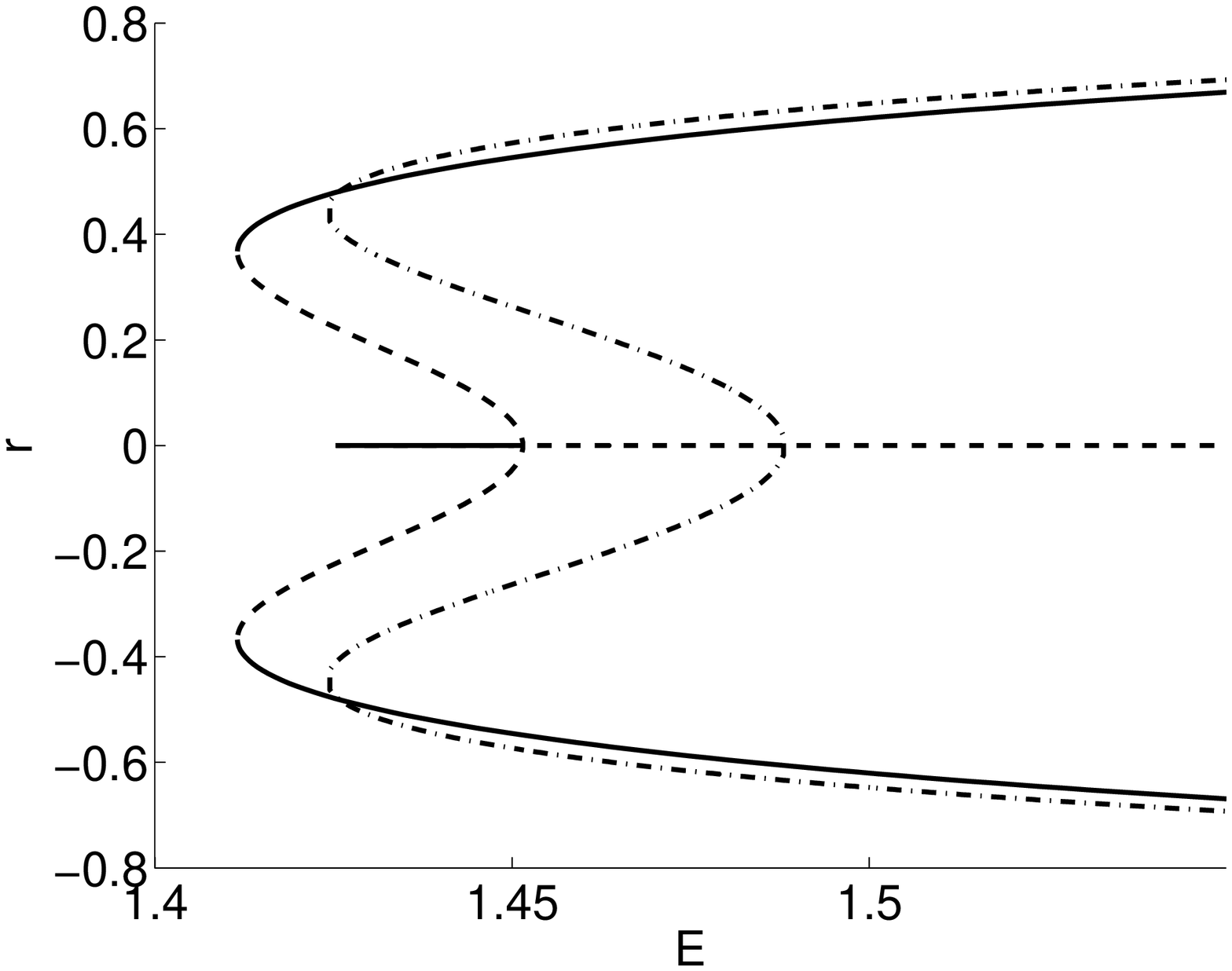}\\[2.0ex]
\includegraphics[width=6.5cm,height=5cm,angle=0,clip]{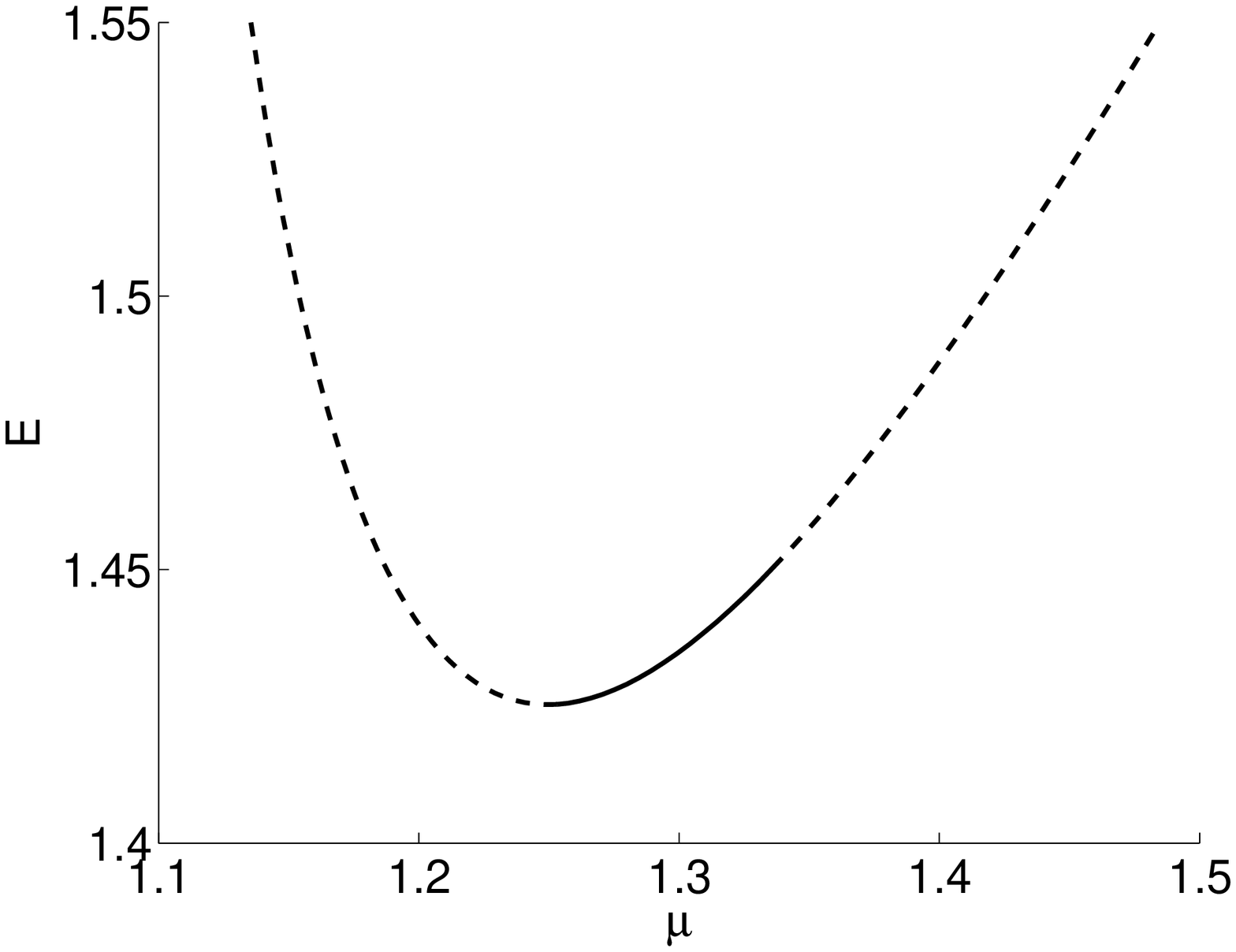}
\caption{The top panel shows the bifurcation diagram in the 2D
model for $\protect\epsilon =0.25$, in the same way as the 1D
diagram is shown in Fig.~\protect\ref{Branches}. The bottom panel displays the
dependence of the solution's squared 
norm, $E$, upon the chemical potential, $\protect%
\mu $, for the symmetric solutions. Unlike the 1D case, there are
two different symmetric solutions for many values of $E$,
resulting in both stable and unstable solutions for norms below
the value at which the symmetric and asymmetric solution branches
intersect.} \label{Branches2D}
\end{figure}

\begin{figure*}[t]
 \includegraphics[width=5.75cm,height=4.5cm,angle=0,clip]{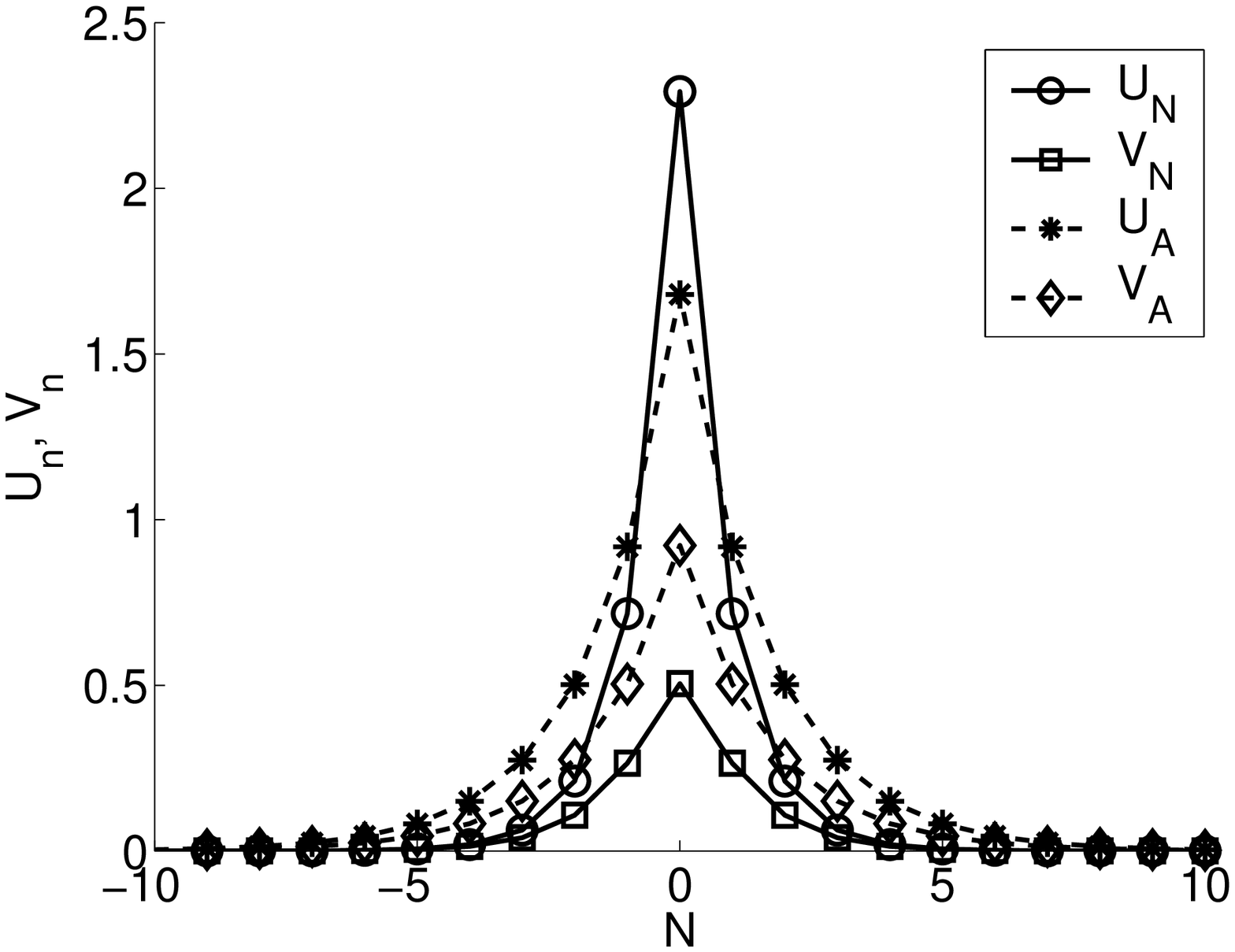} %
~\includegraphics[width=5.75cm,height=4.5cm,angle=0,clip]{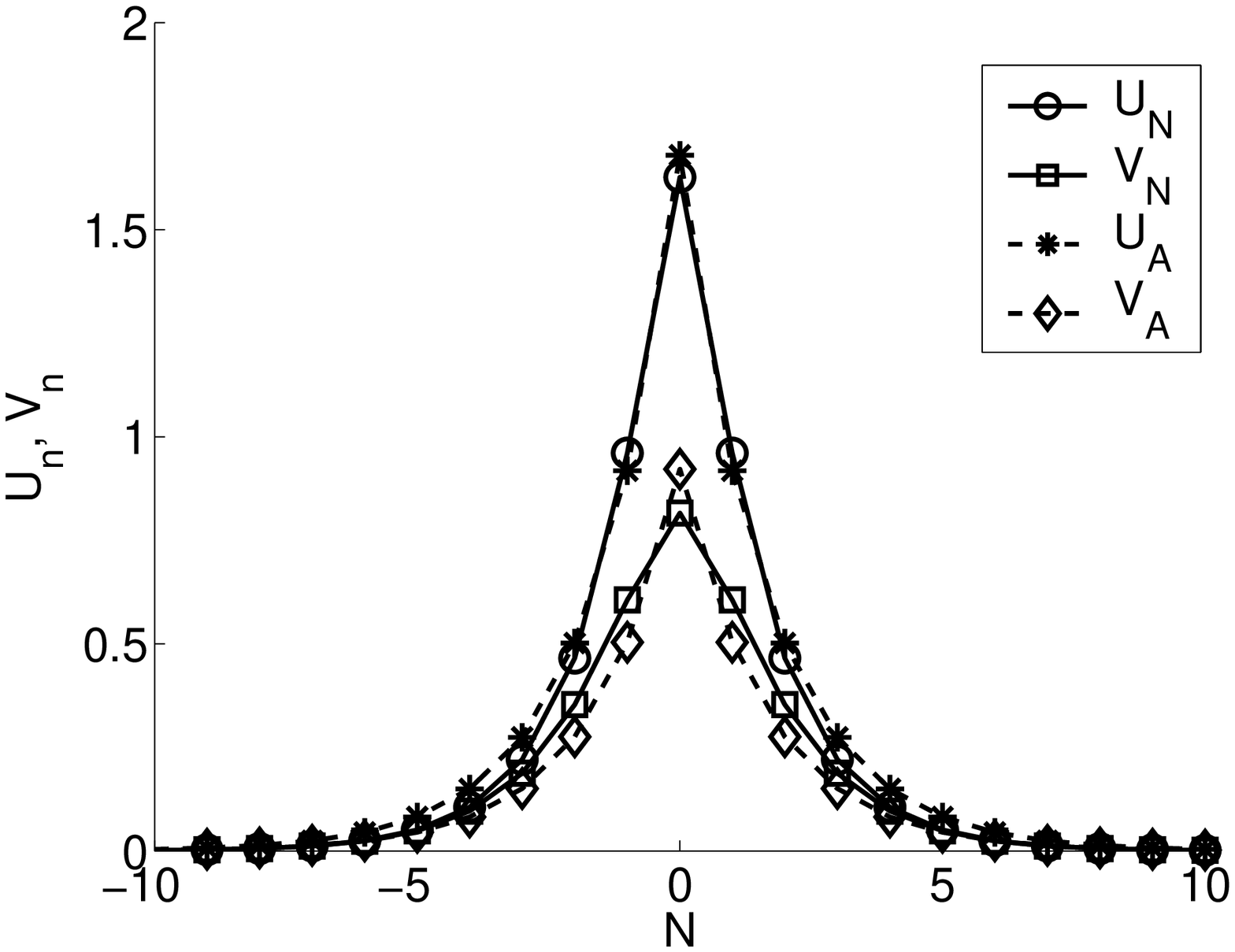} %
~\includegraphics[width=5.75cm,height=4.5cm,angle=0,clip]{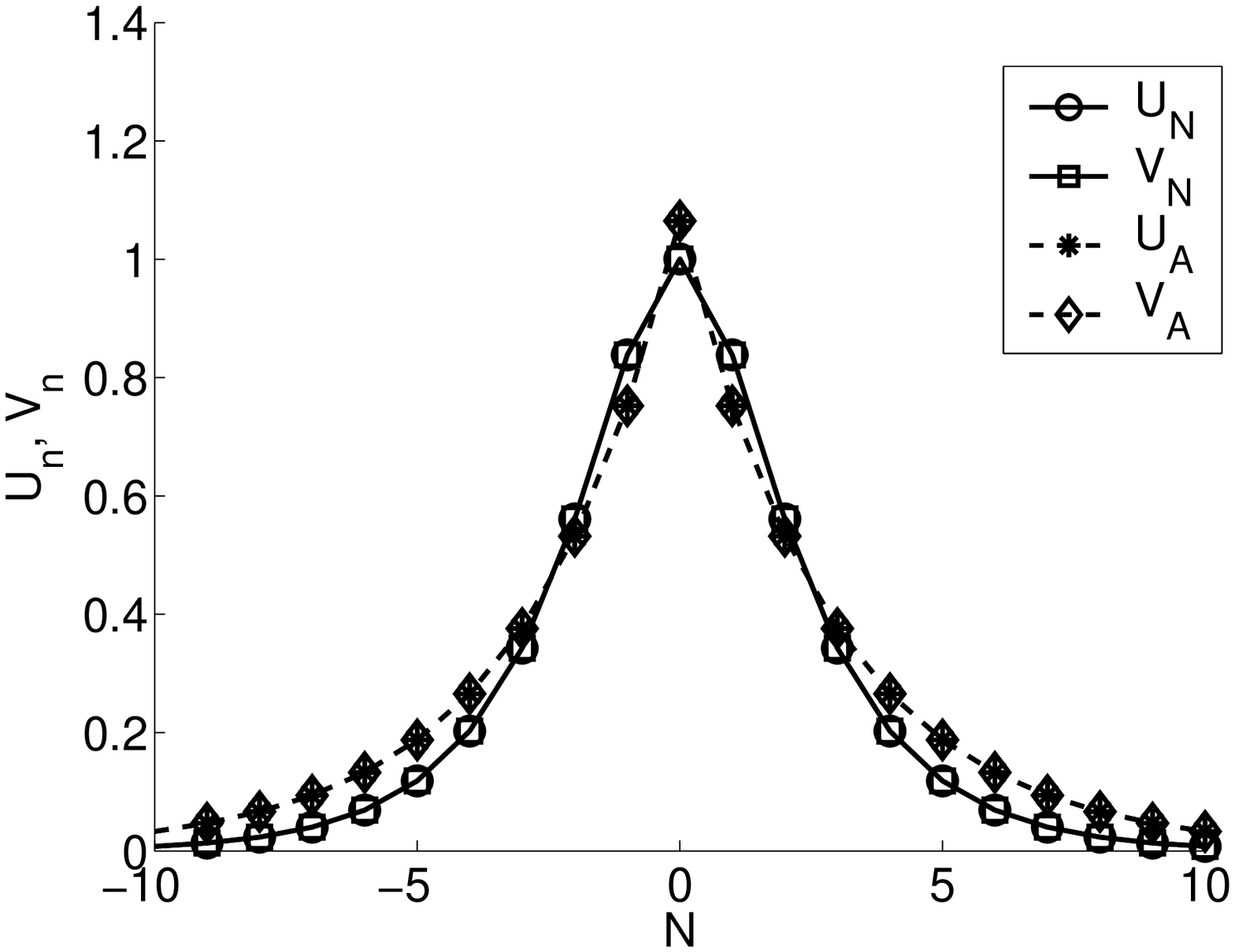}\\[2.0ex]
 \includegraphics[width=5.75cm,height=4.5cm,angle=0,clip]{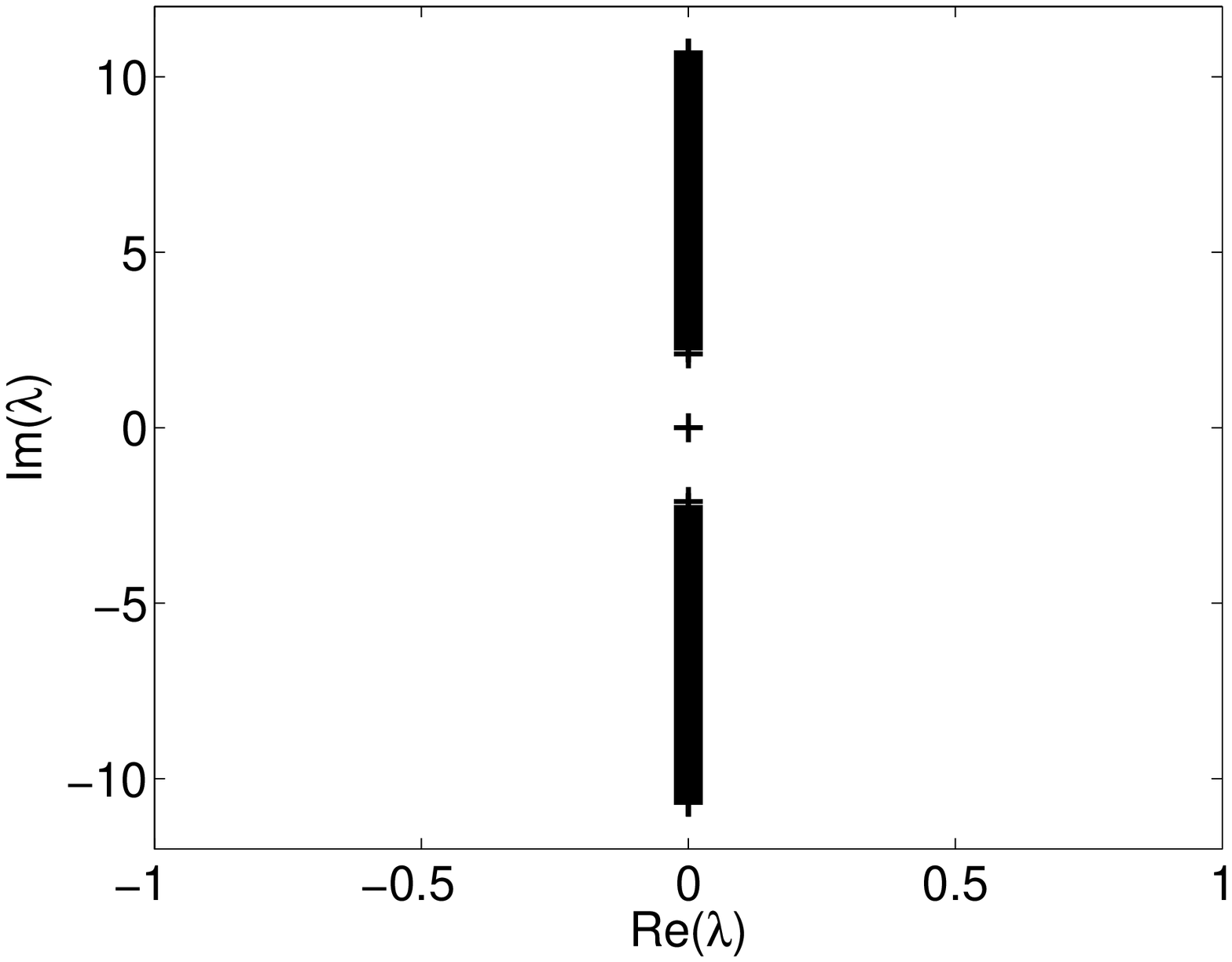} %
~\includegraphics[width=5.75cm,height=4.5cm,angle=0,clip]{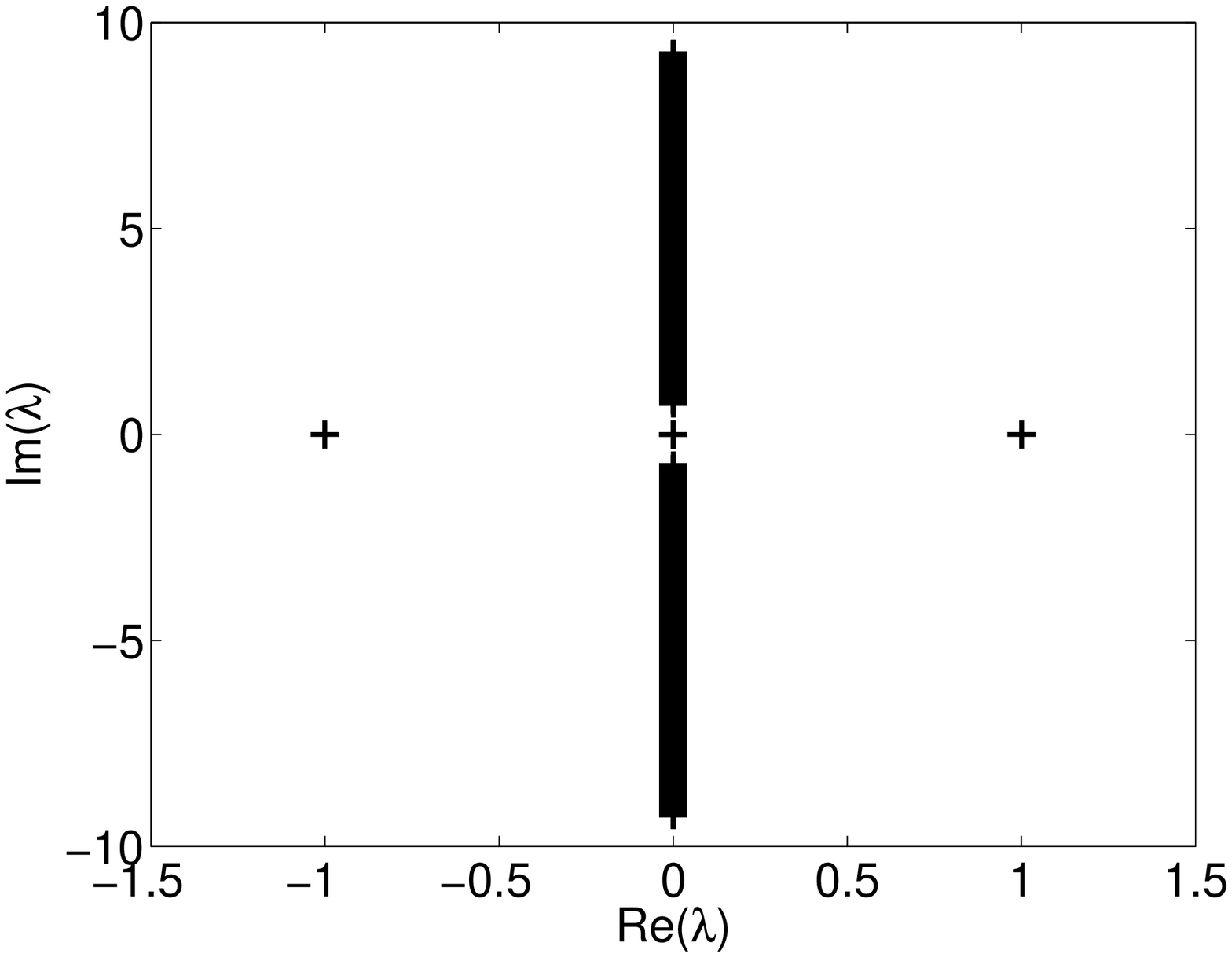} %
~\includegraphics[width=5.75cm,height=4.5cm,angle=0,clip]{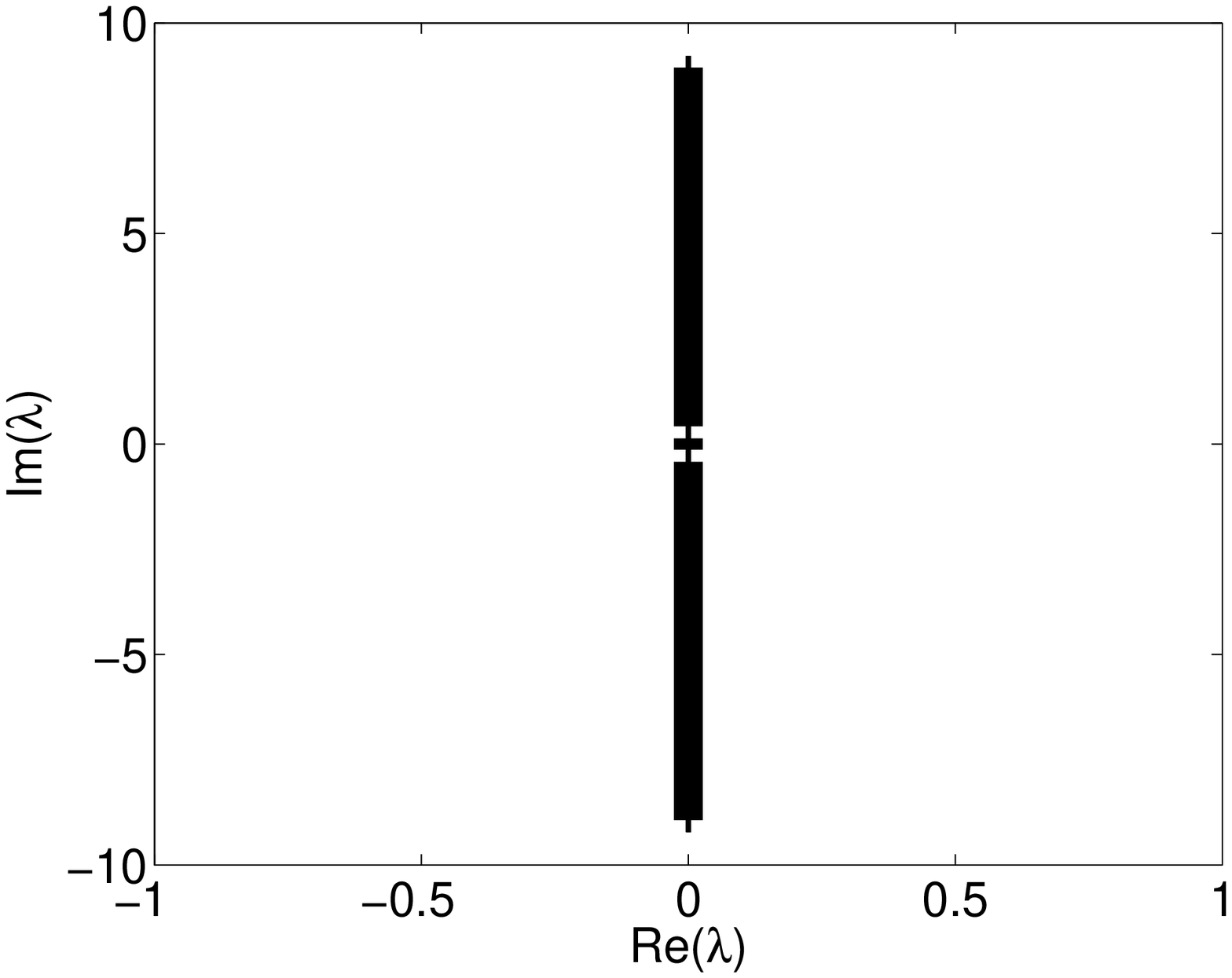}
\caption{Plots of solutions from the branch in Fig.~\protect\ref{Branches}
for $E=3.4$. The top-row figures show the
solution profiles found by means of the numerical ($U_{N},V_{N}$)
and variational (``analytical", $U_{A},V_{A}$) methods. The bottom
row plots illustrate the stability eigenvalues for the numerical solution.
The first column presents a stable
stationary asymmetric solution from the outer (upper) branch  in
Fig.~\protect\ref{Branches},
the second column is an unstable asymmetric solution of the inner
branch, and the last column is taken from the stable part of the family of
symmetric solutions, with $r=0$.}
\label{Solns}
\end{figure*}

\begin{figure*}[t]
 \includegraphics[width=5.75cm,height=4.5cm,angle=0,clip]{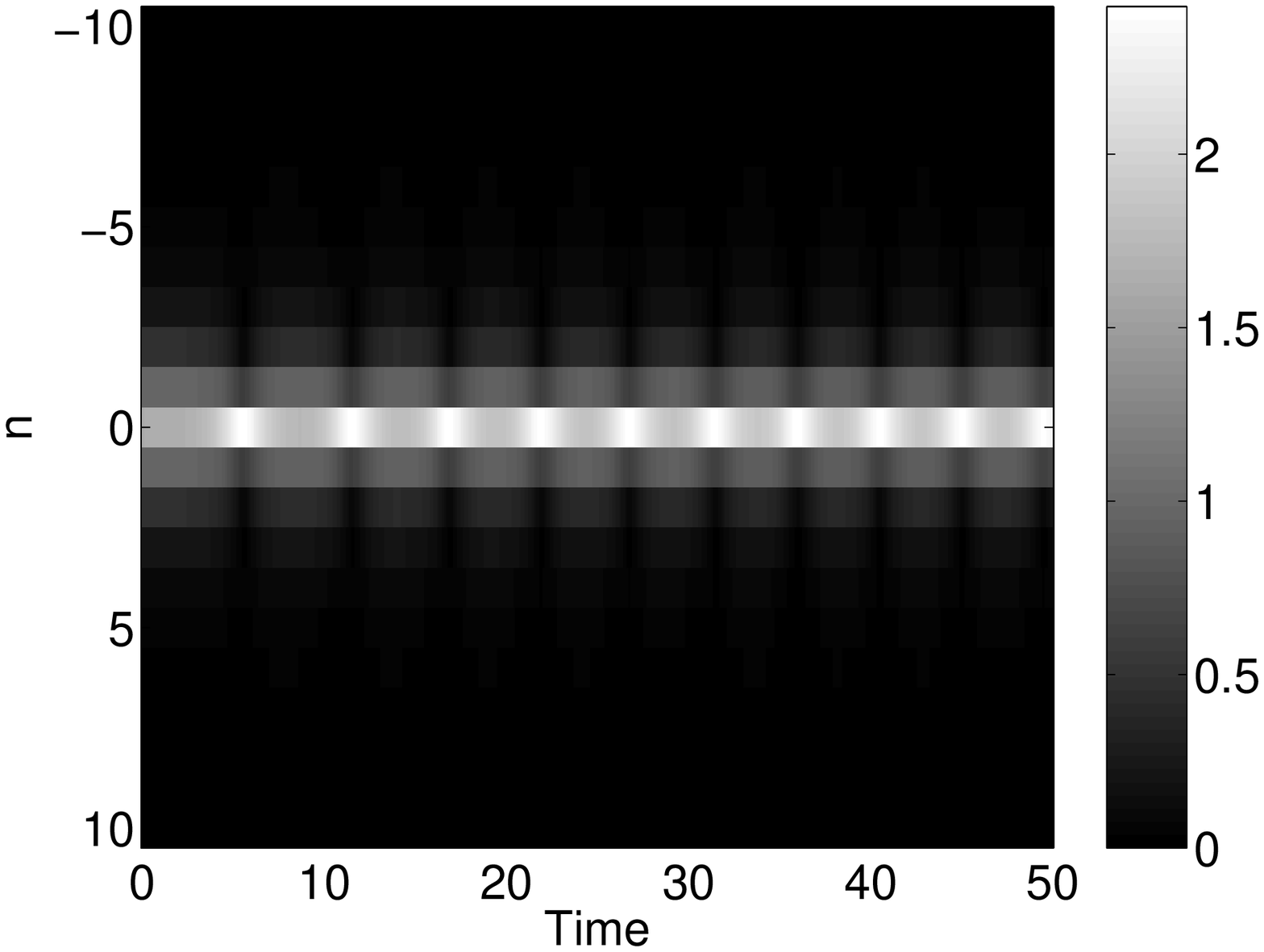} %
~\includegraphics[width=5.75cm,height=4.5cm,angle=0,clip]{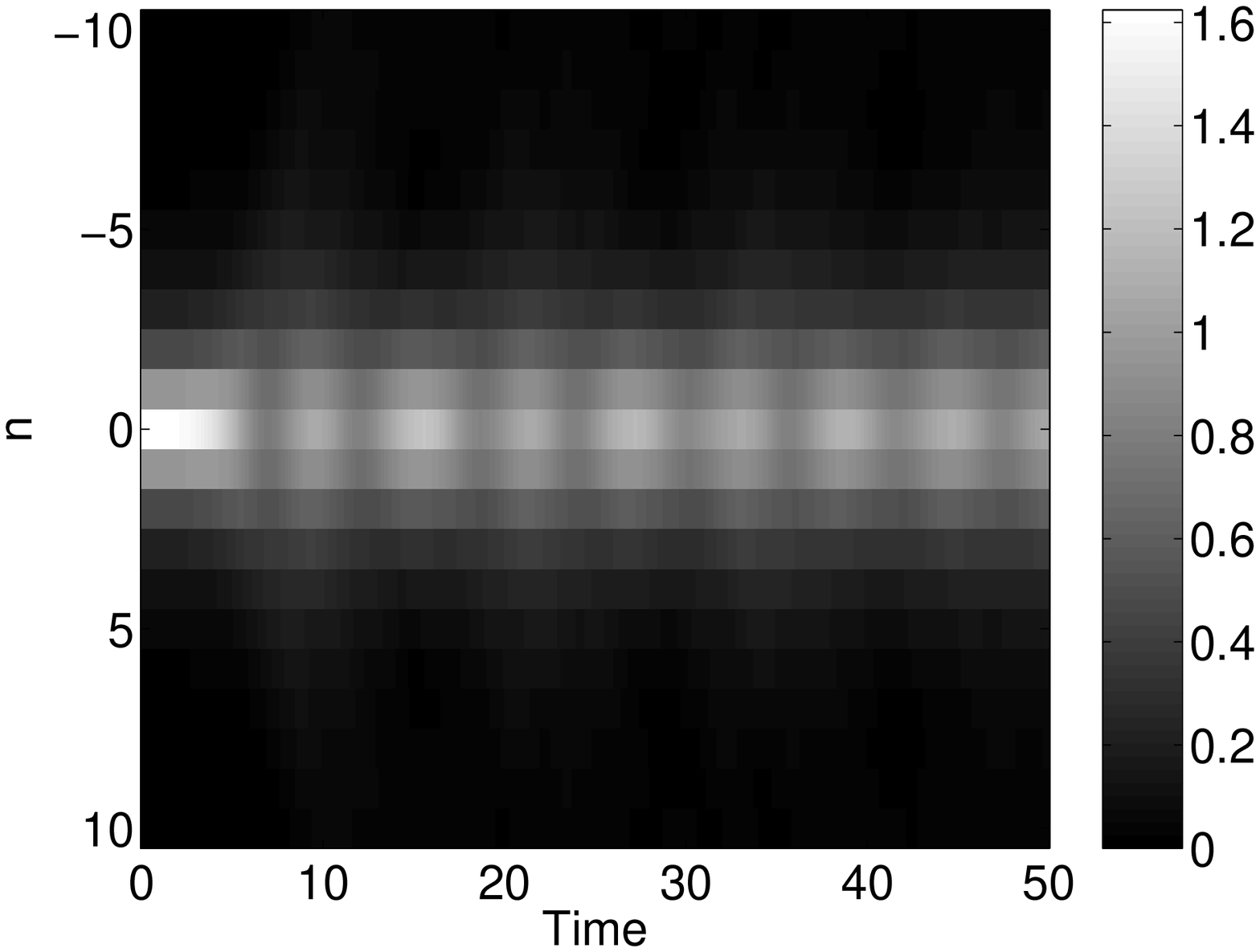} %
~\includegraphics[width=5.75cm,height=4.5cm,angle=0,clip]{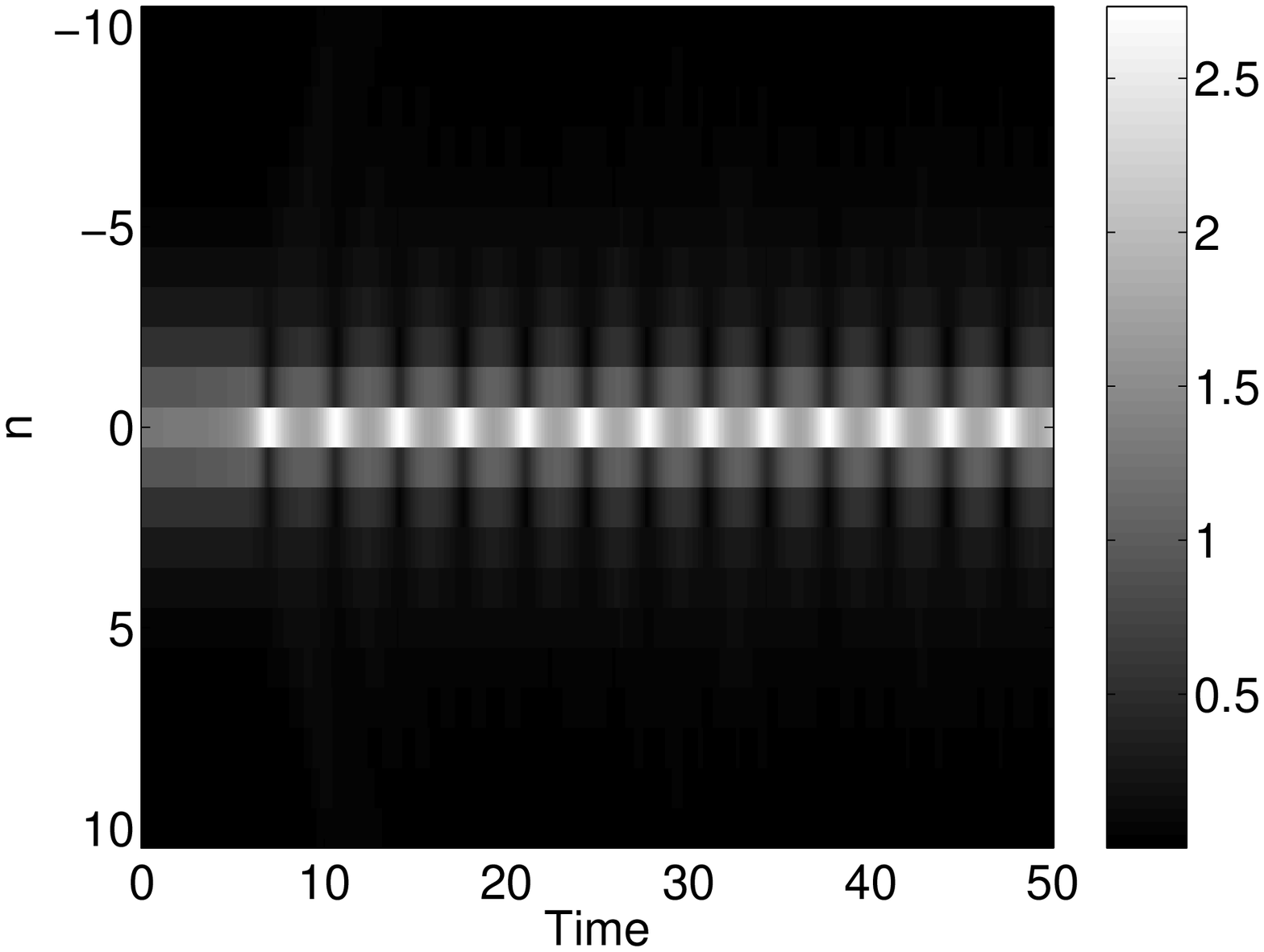}\\[2.ex]
 \includegraphics[width=5.75cm,height=4.5cm,angle=0,clip]{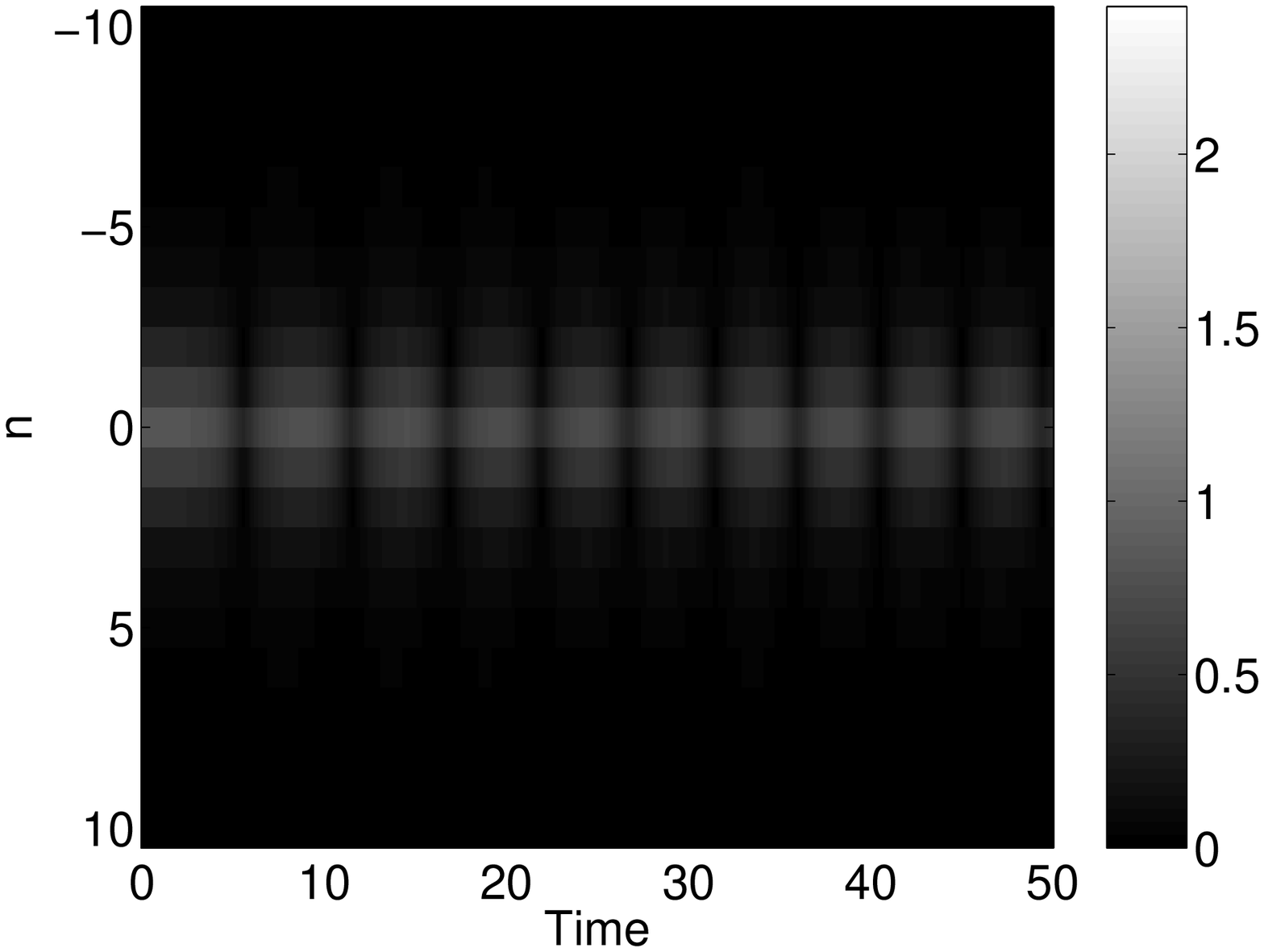} %
~\includegraphics[width=5.75cm,height=4.5cm,angle=0,clip]{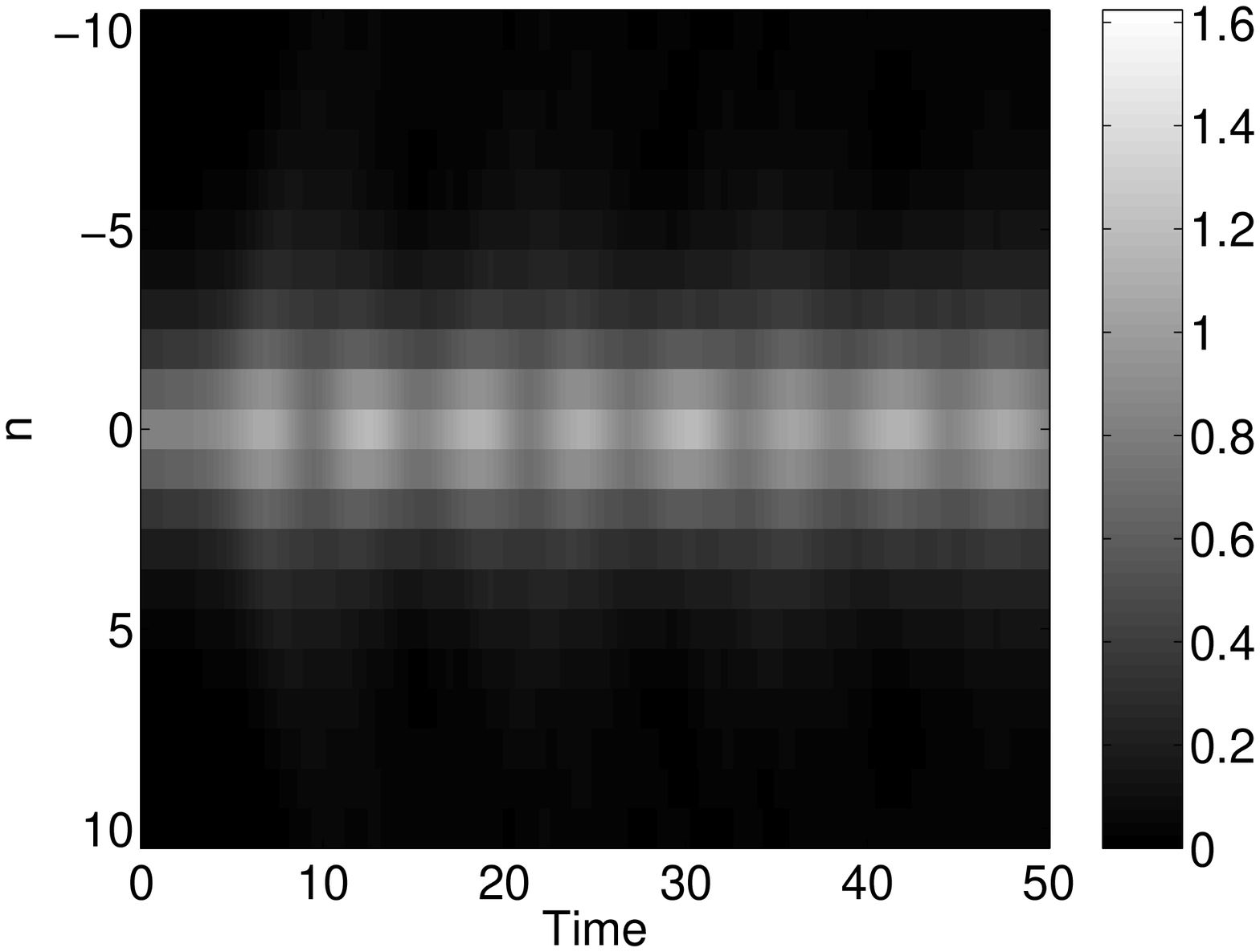} %
~\includegraphics[width=5.75cm,height=4.5cm,angle=0,clip]{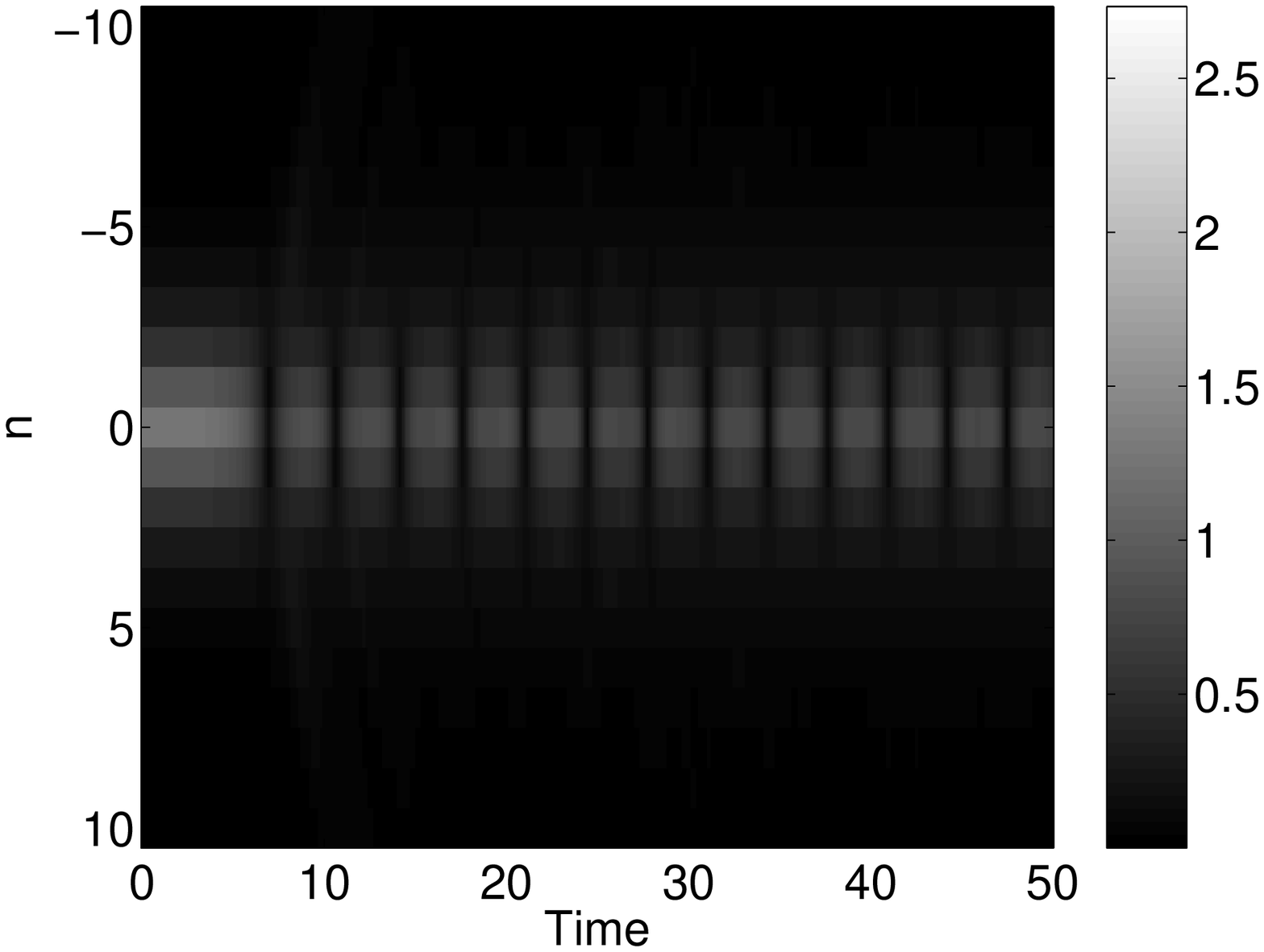}\\[2.ex]
 \includegraphics[width=5.75cm,height=4.5cm,angle=0,clip]{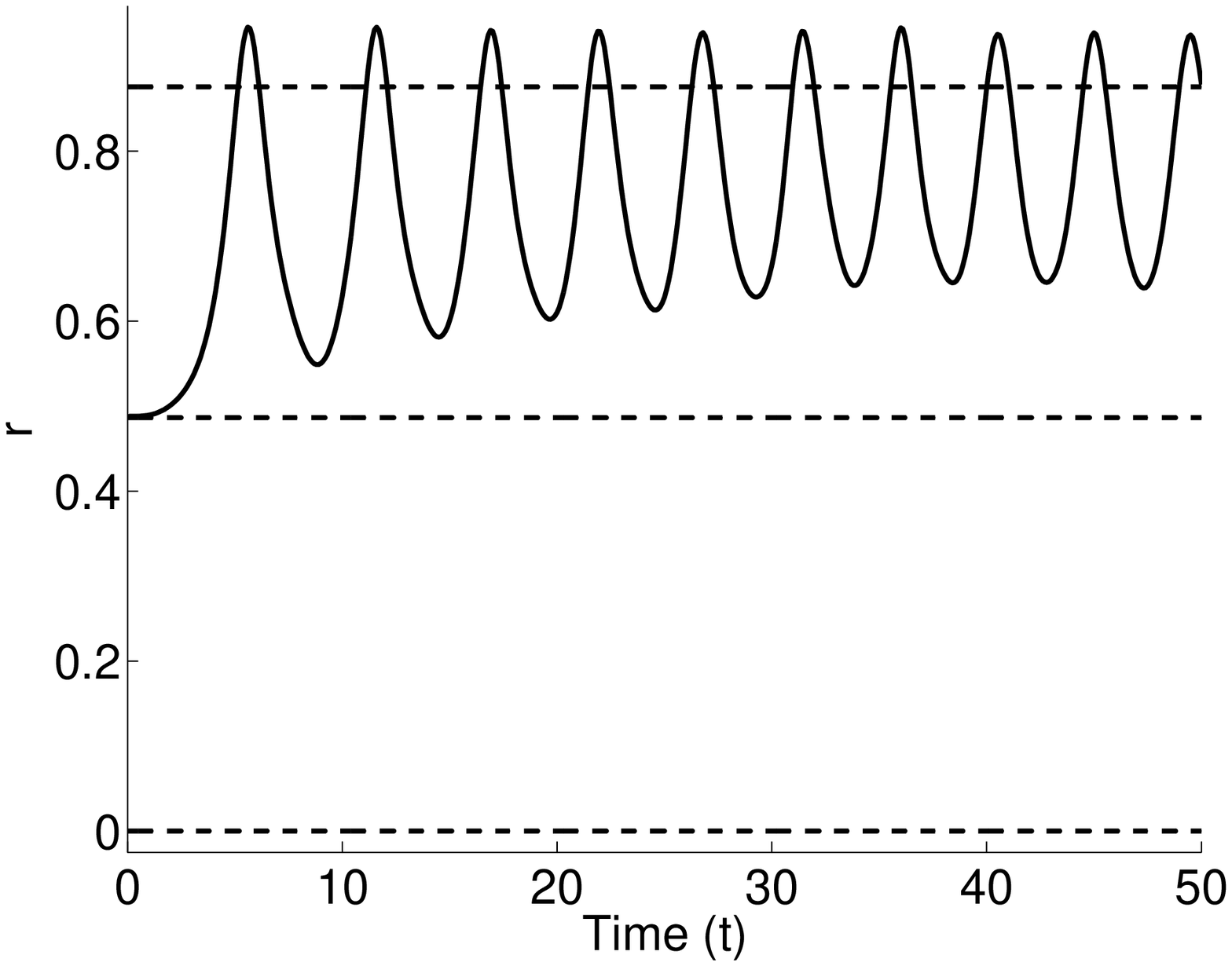} %
~\includegraphics[width=5.75cm,height=4.5cm,angle=0,clip]{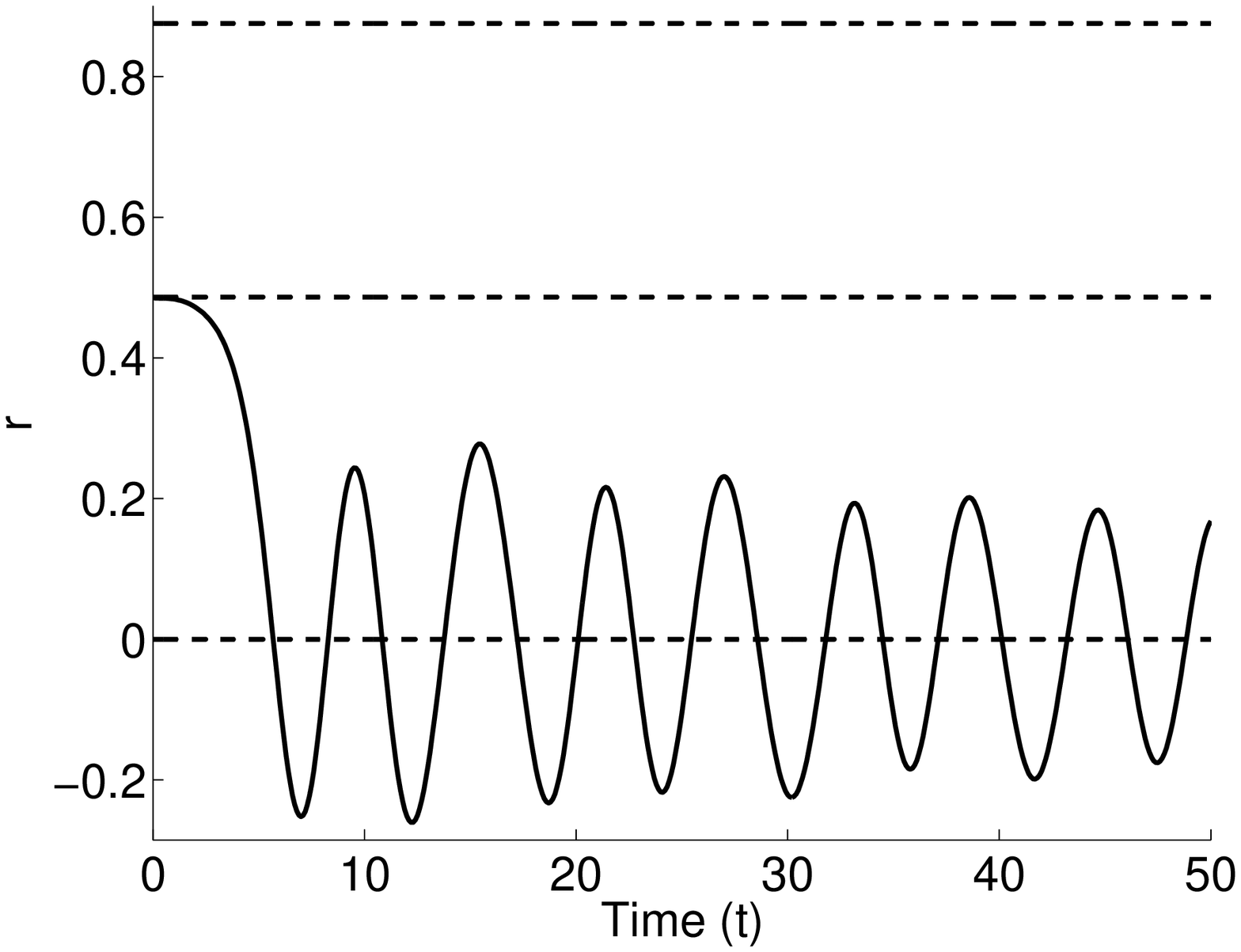} %
~\includegraphics[width=5.75cm,height=4.5cm,angle=0,clip]{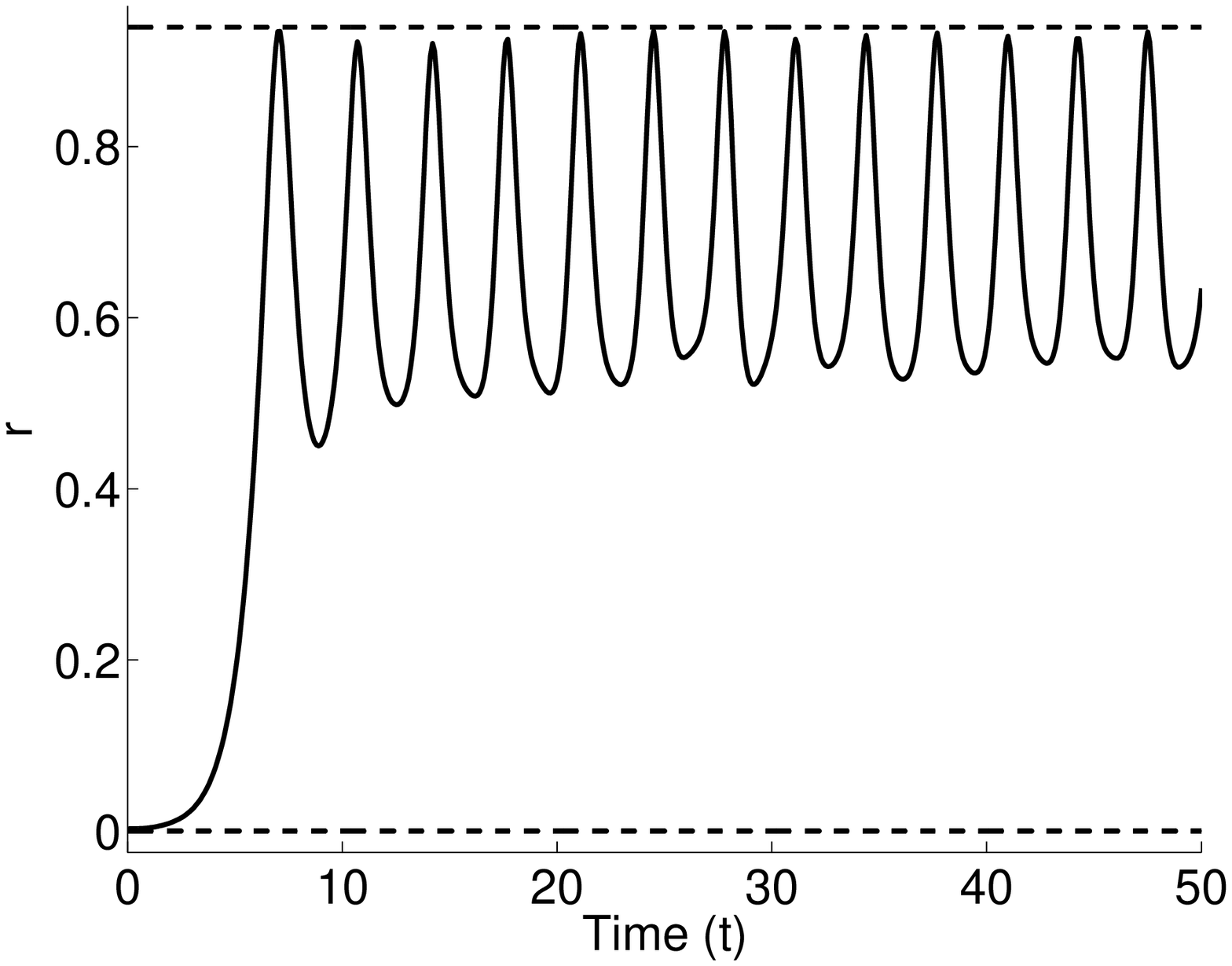}
\caption{Time-evolution plots for perturbed one-dimensional solutions from
Fig.~\protect\ref{Branches}.
The top (middle) row depicts the space-time contour plot evolution of
$|U|^2$ ($|V|^2$).
The first two columns are for two different
perturbations of the solution shown in the second column of Fig.~\protect
\ref{Solns}: the first column is generated by a perturbation which pushes
the solution towards the upper stable asymmetric 
solution branch of the bifurcation diagram,
while, in the second column, the perturbation pushes the solution towards
the $r=0$ solution. The last column is for a perturbation of the unstable
symmetric solution (on the $r=0$ curve) at $E=4.1$, which pushes the
solution towards the upper stable solution branch. The bottom row of figures
displays the value of $r$ as the solutions evolve in time, shown by the solid
line, and the constant value of $r$ for the steady states at the same energy
level (dashed horizontal lines).}
\label{TimeEvo}
\end{figure*}

\begin{figure*}[ht]
 \includegraphics[width=5.75cm,height=4.5cm,angle=0,clip]{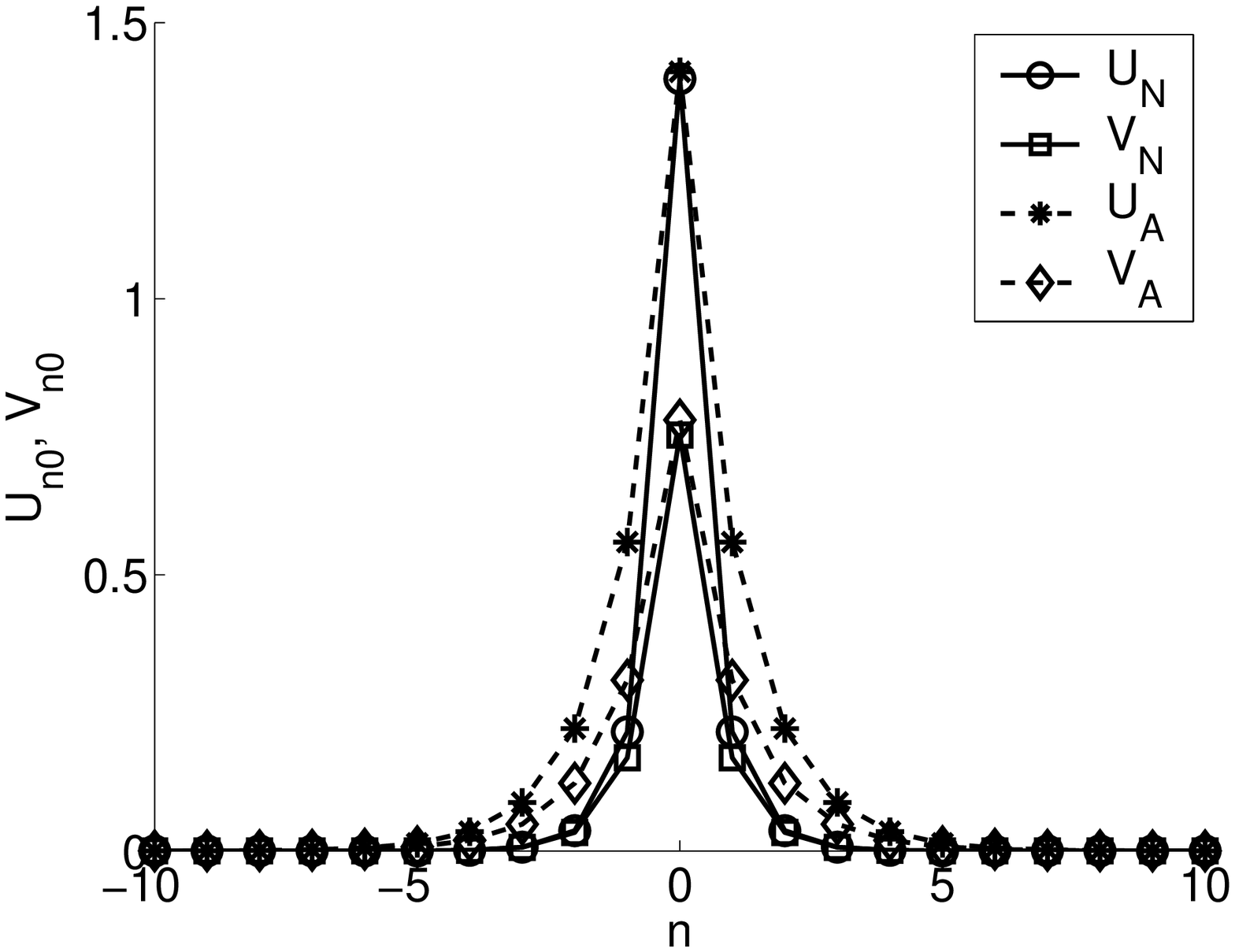} %
~\includegraphics[width=5.75cm,height=4.5cm,angle=0,clip]{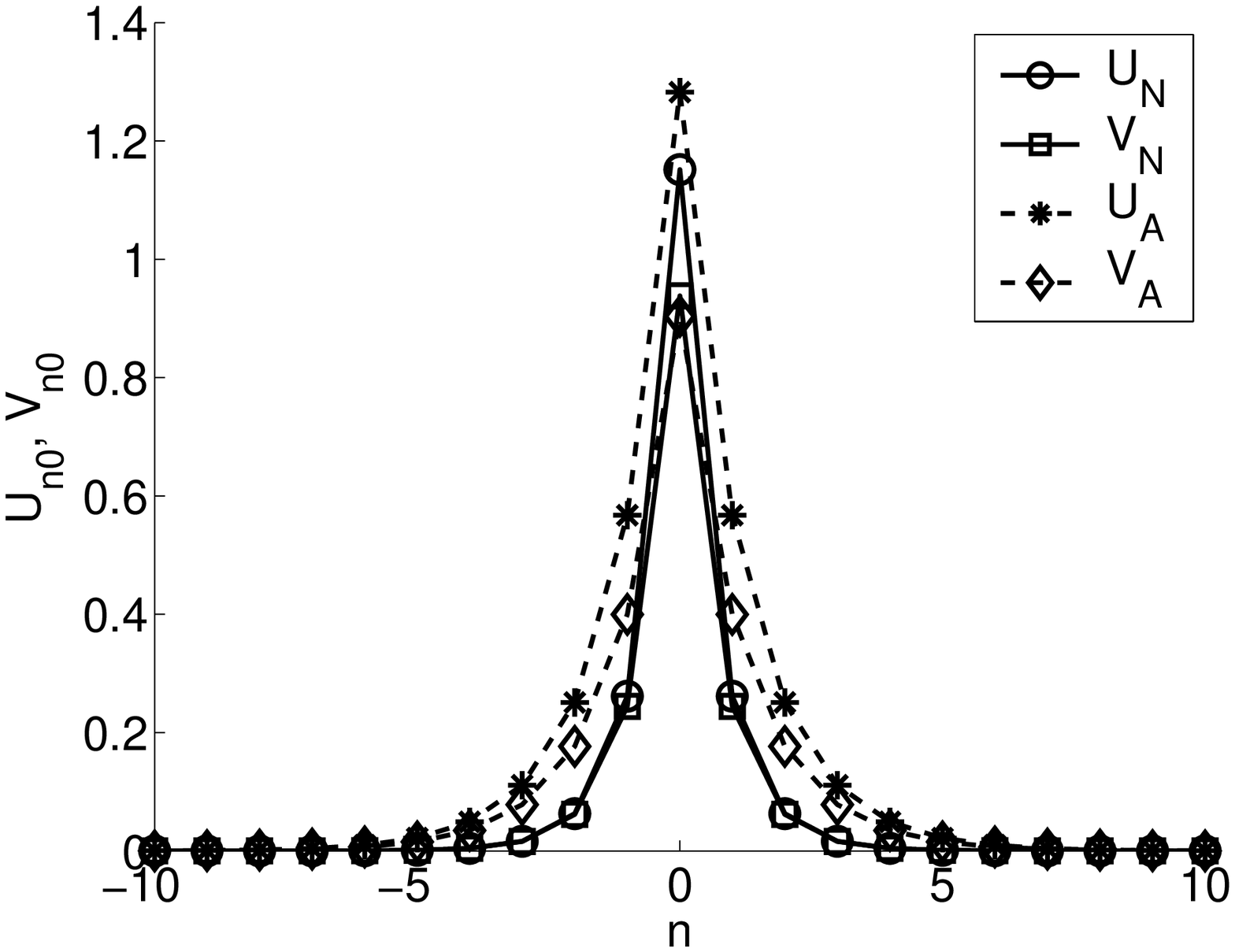}\\[2.0ex]
 \includegraphics[width=5.75cm,height=4.5cm,angle=0,clip]{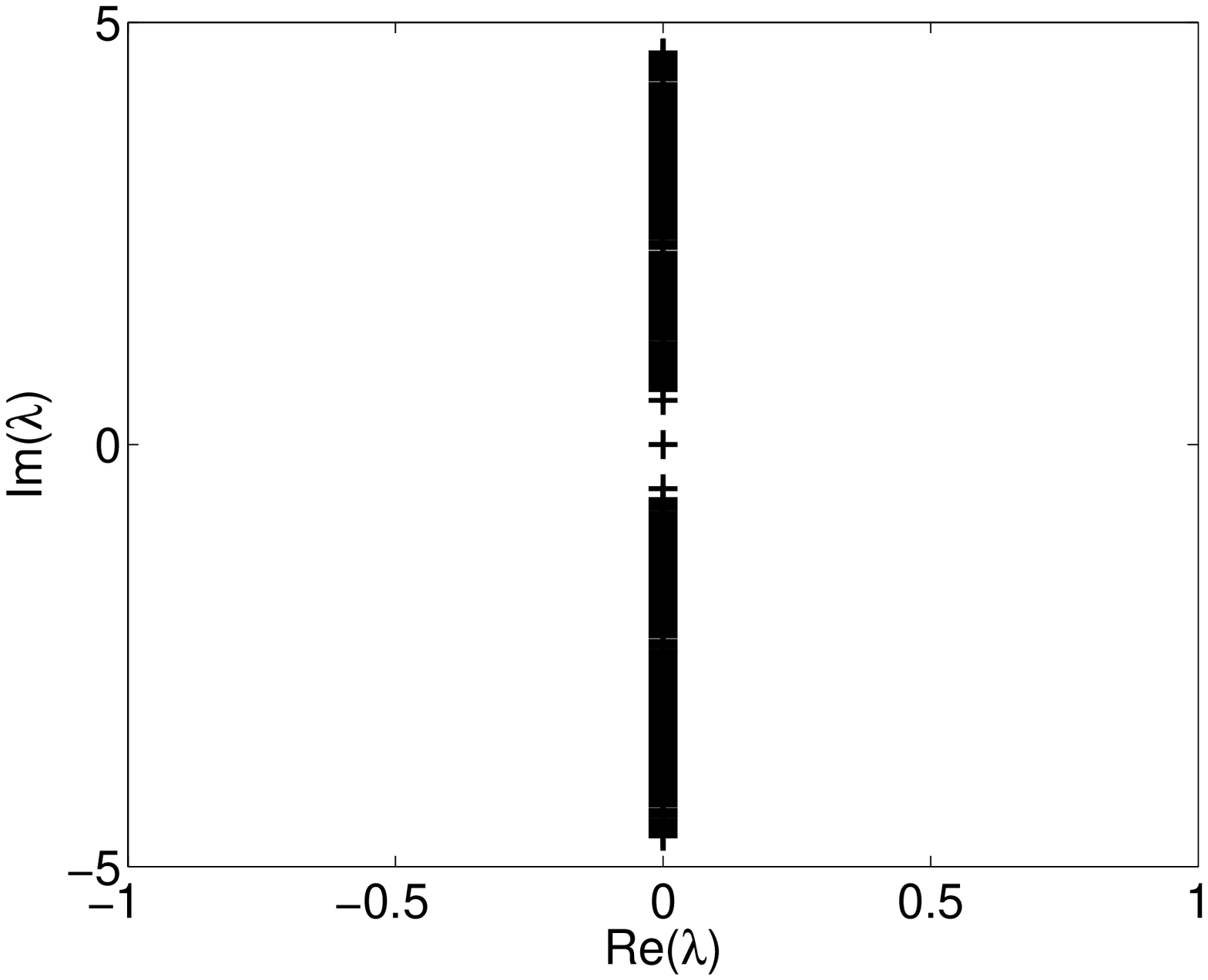} %
~\includegraphics[width=5.75cm,height=4.5cm,angle=0,clip]{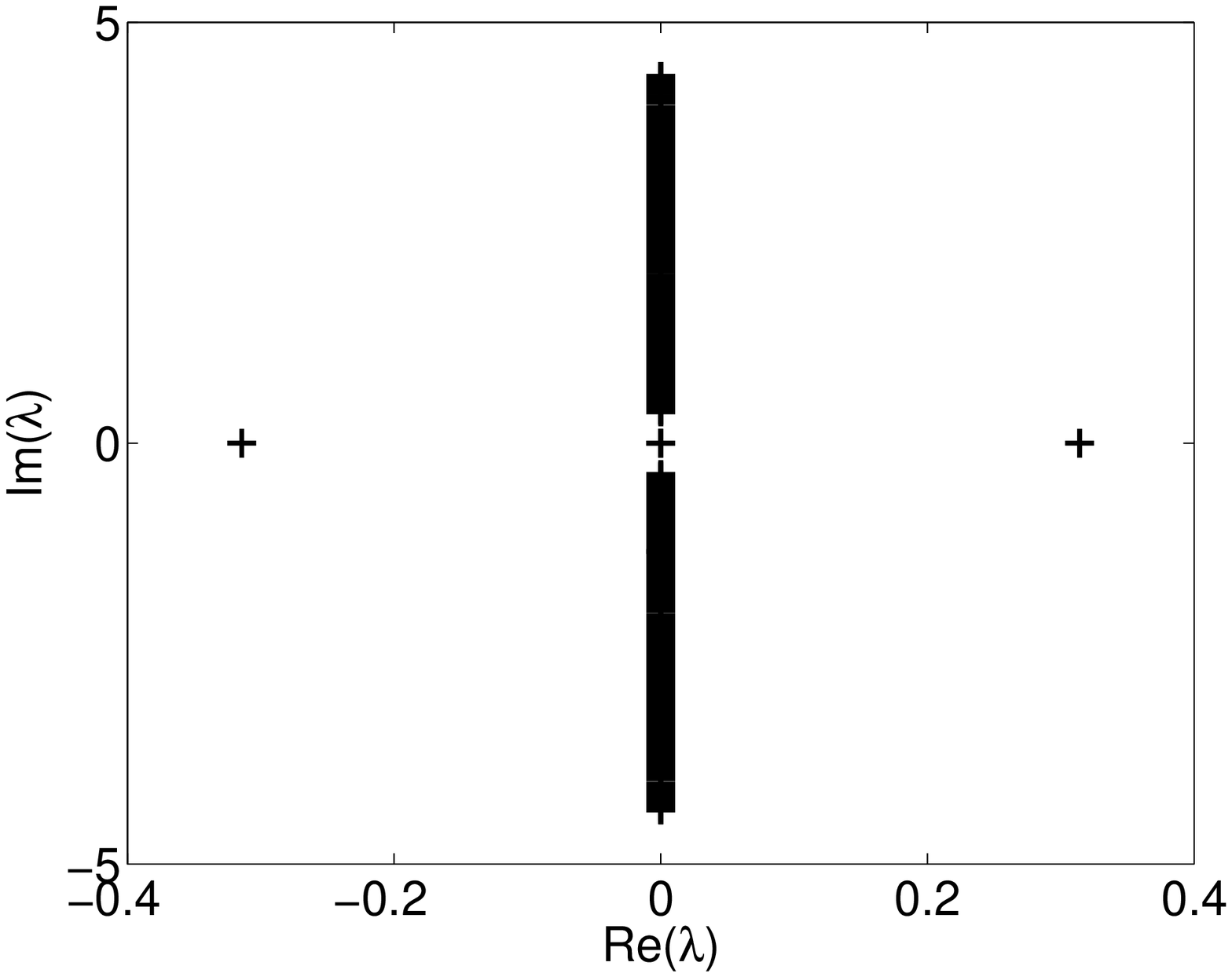}
\caption{Cross-section plots of the asymmetric solutions belonging to
corresponding curve in Fig.~\protect\ref{Branches2D}, for $E=1.435$. The
top-row figures show the solutions found by means of the numerical ($%
U_{N},V_{N}$) and variational (``analytical", $U_{A},V_{A}$)
methods, and the bottom row plots illustrate the stability eigenvalues for the
numerical solution. As can be seen, the first and second columns
represent, respectively, stable and unstable solutions belonging
to the upper (outer) and inner curves (asymmetric branches)
of the bifurcation diagram, respectively.}
\label{Solns2D1}
\end{figure*}

\begin{figure*}[ht]
 \includegraphics[width=5.75cm,height=4.5cm,angle=0,clip]{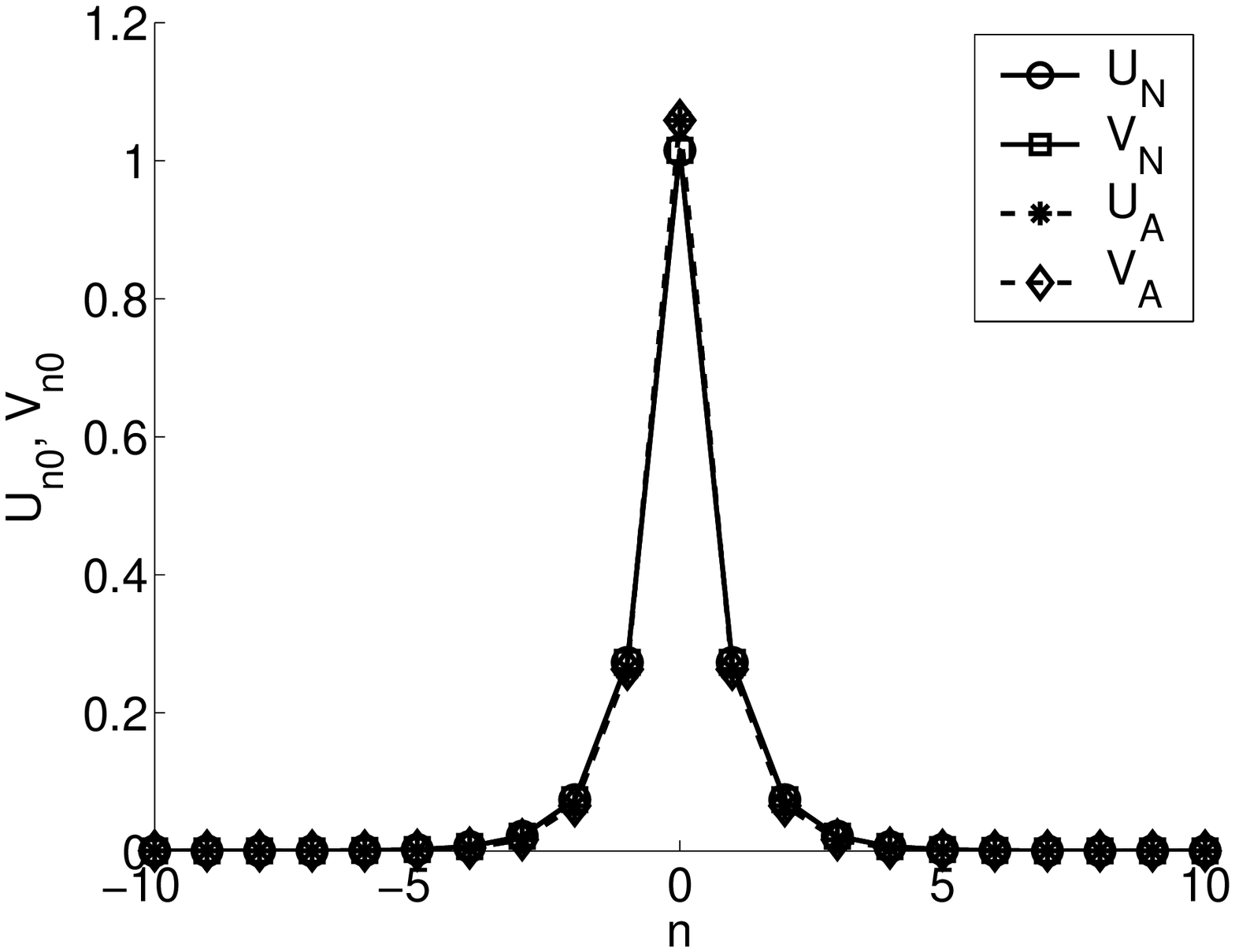} %
~\includegraphics[width=5.75cm,height=4.5cm,angle=0,clip]{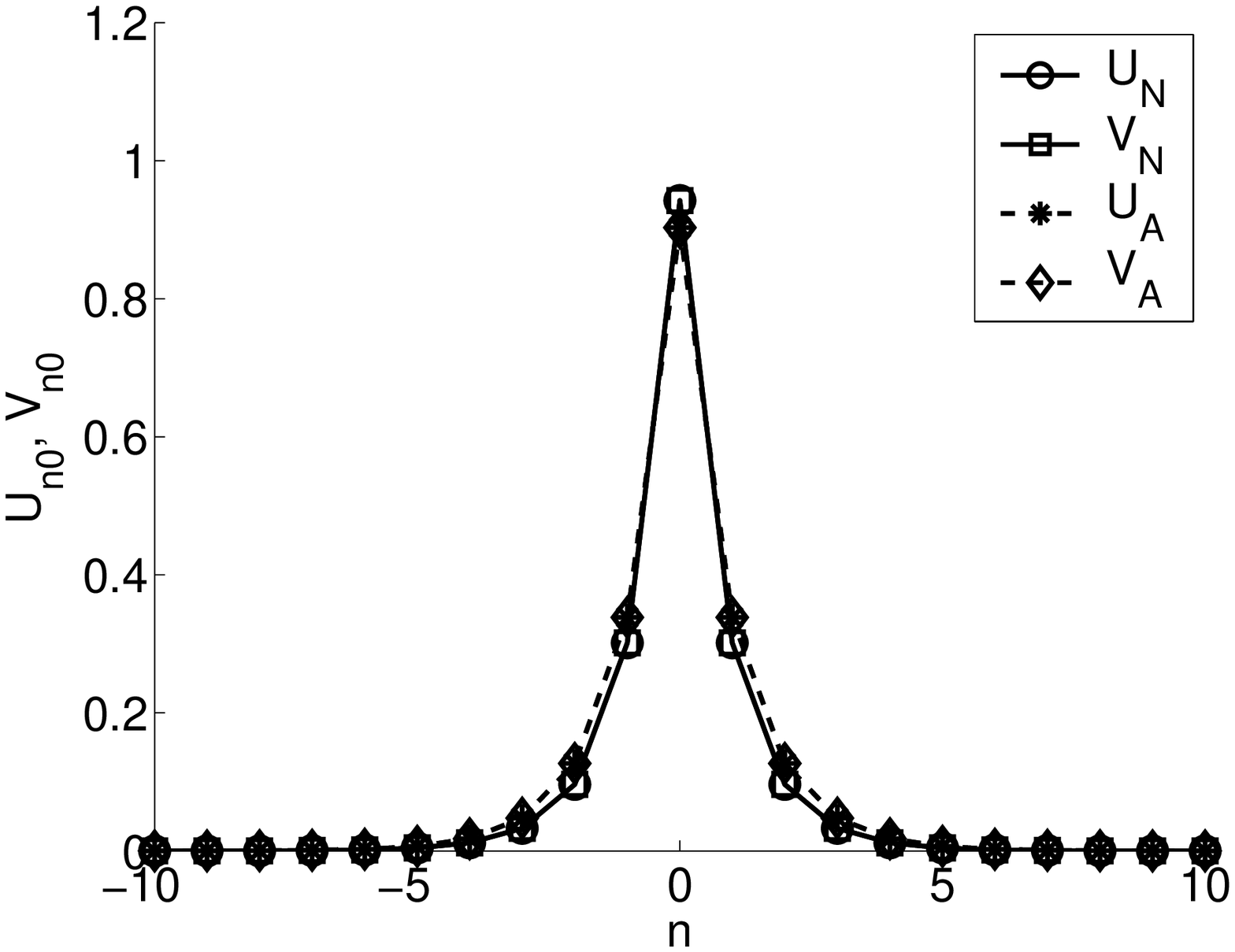}\\[2.0ex]
 \includegraphics[width=5.75cm,height=4.5cm,angle=0,clip]{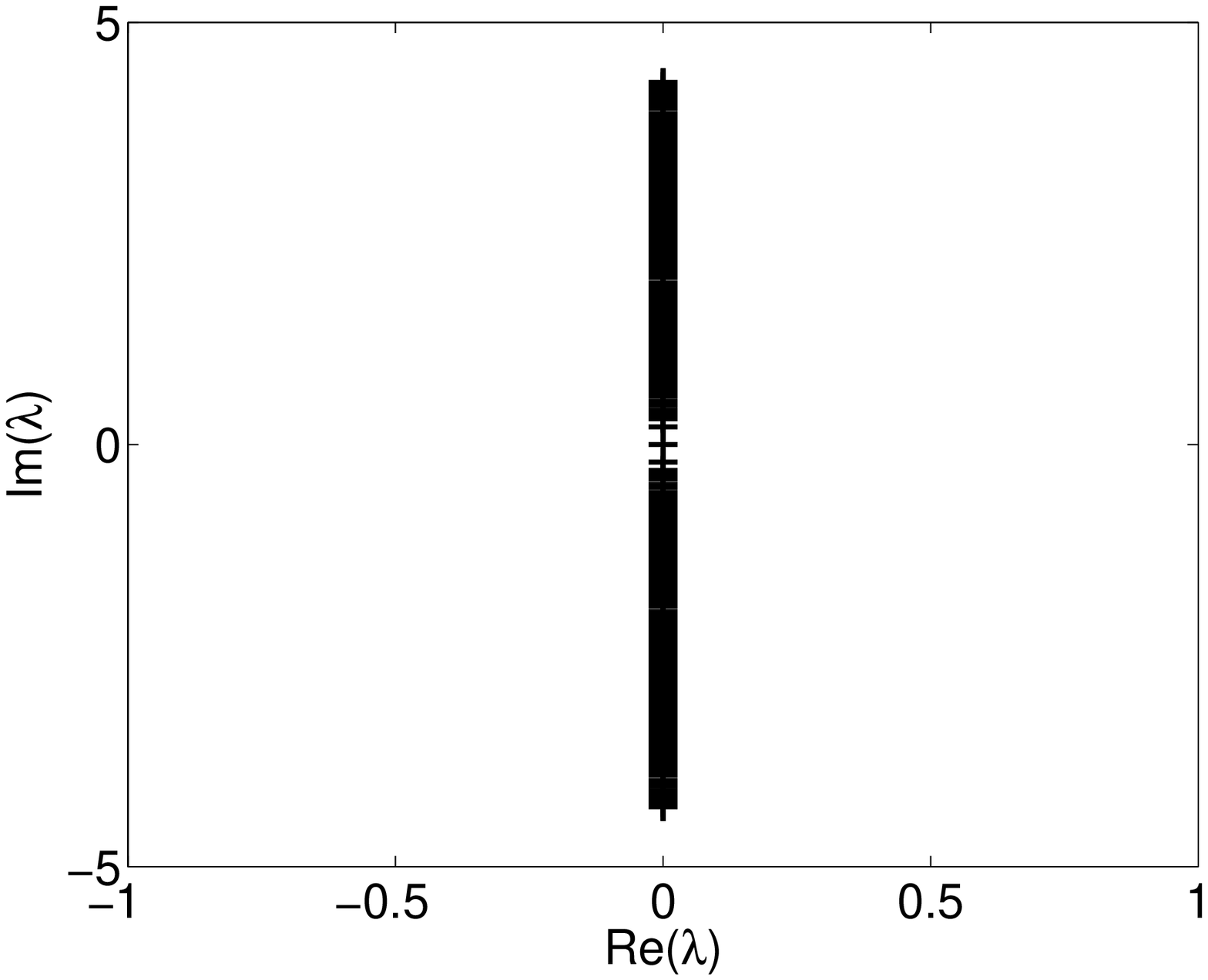} %
~\includegraphics[width=5.75cm,height=4.5cm,angle=0,clip]{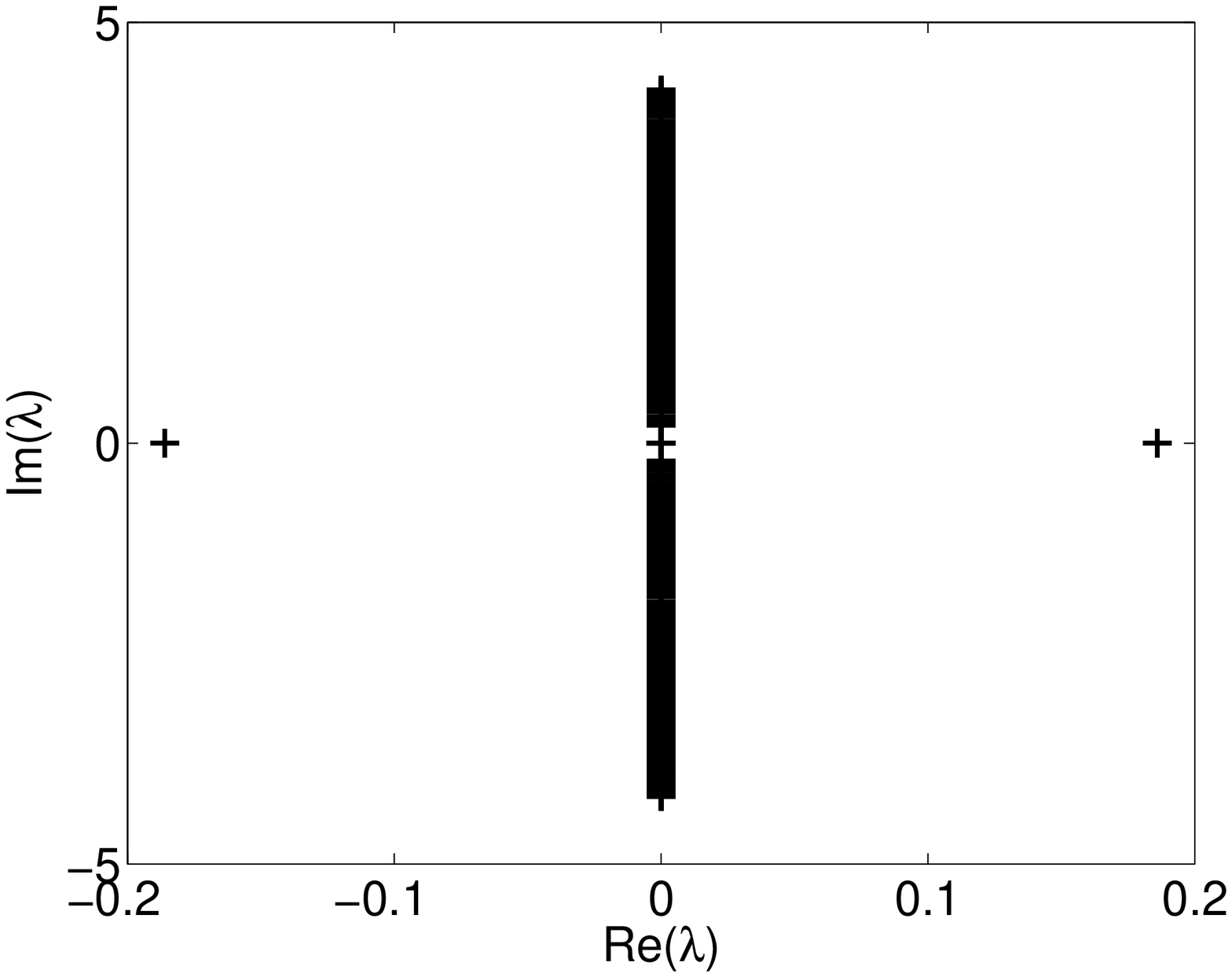}
\caption{Same as Fig.~(\protect\ref{Solns2D1}) for two symmetric solutions
found at $E=1.435$. }
\label{Solns2D2}
\end{figure*}

\begin{figure*}[ht]
\begin{tabular}{cc}
 \includegraphics[width=5.75cm,height=4.5cm,angle=0,clip]{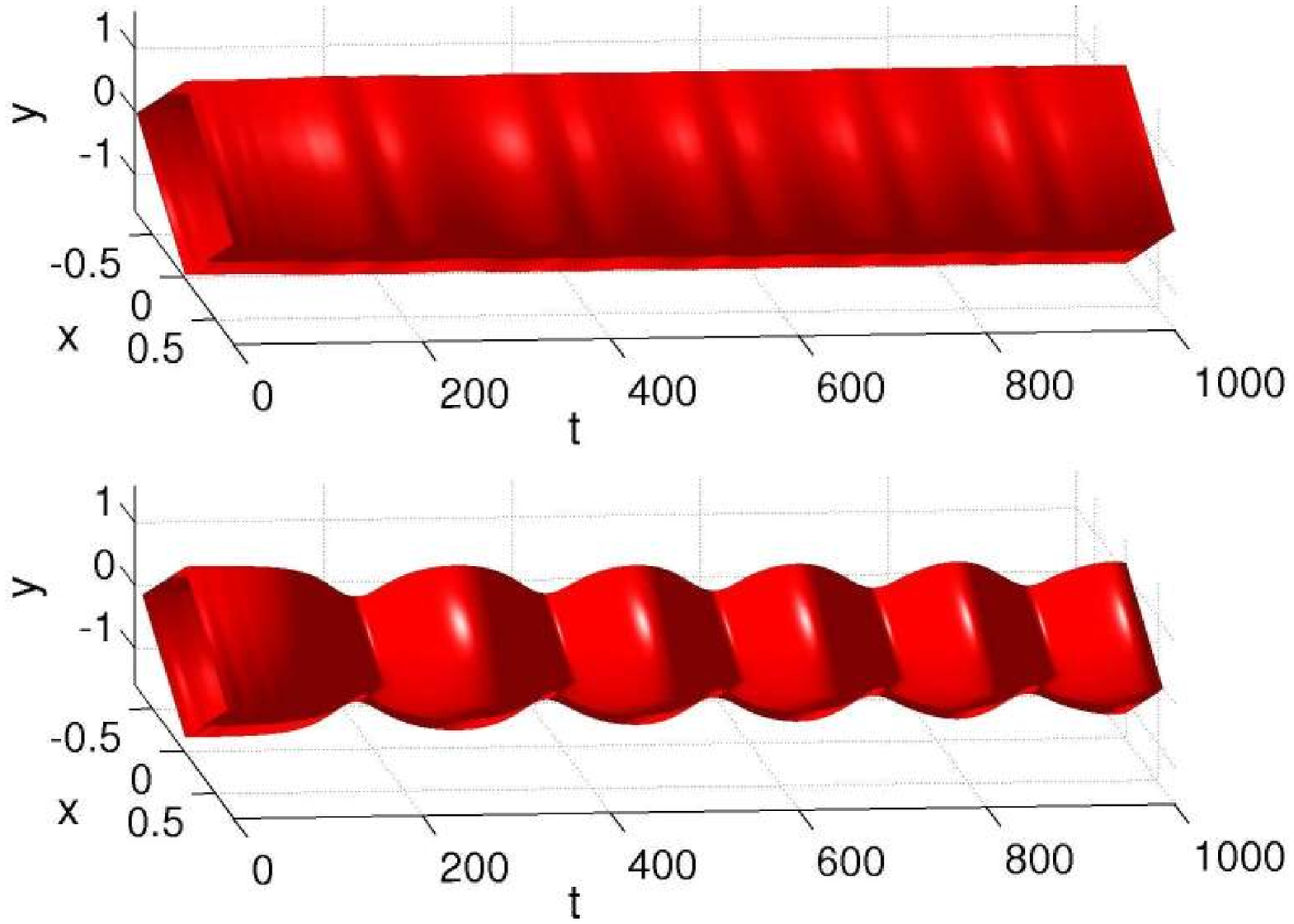}&
~\includegraphics[width=5.75cm,height=4.5cm,angle=0,clip]{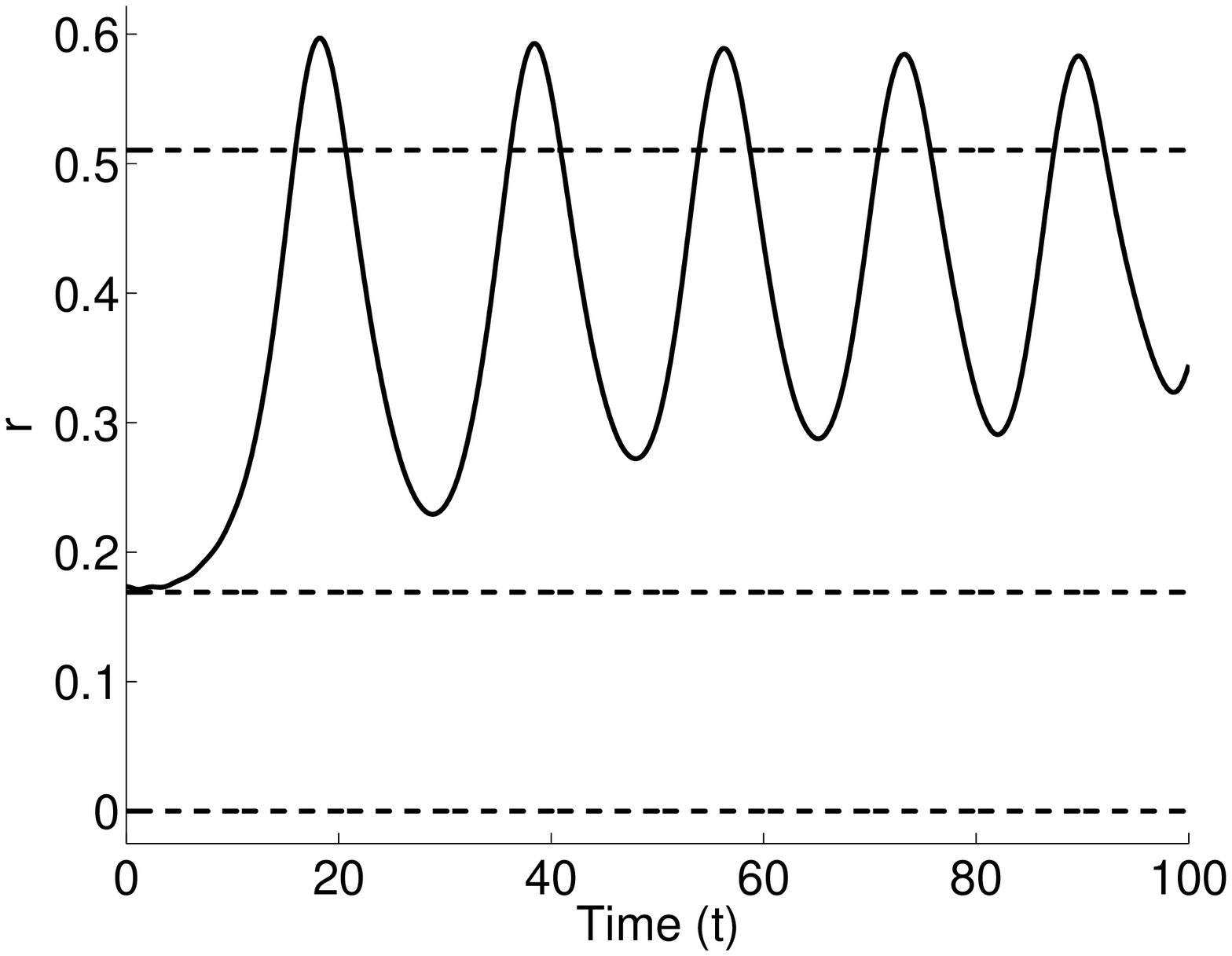}\\[2.0ex]
 \includegraphics[width=5.75cm,height=4.5cm,angle=0,clip]{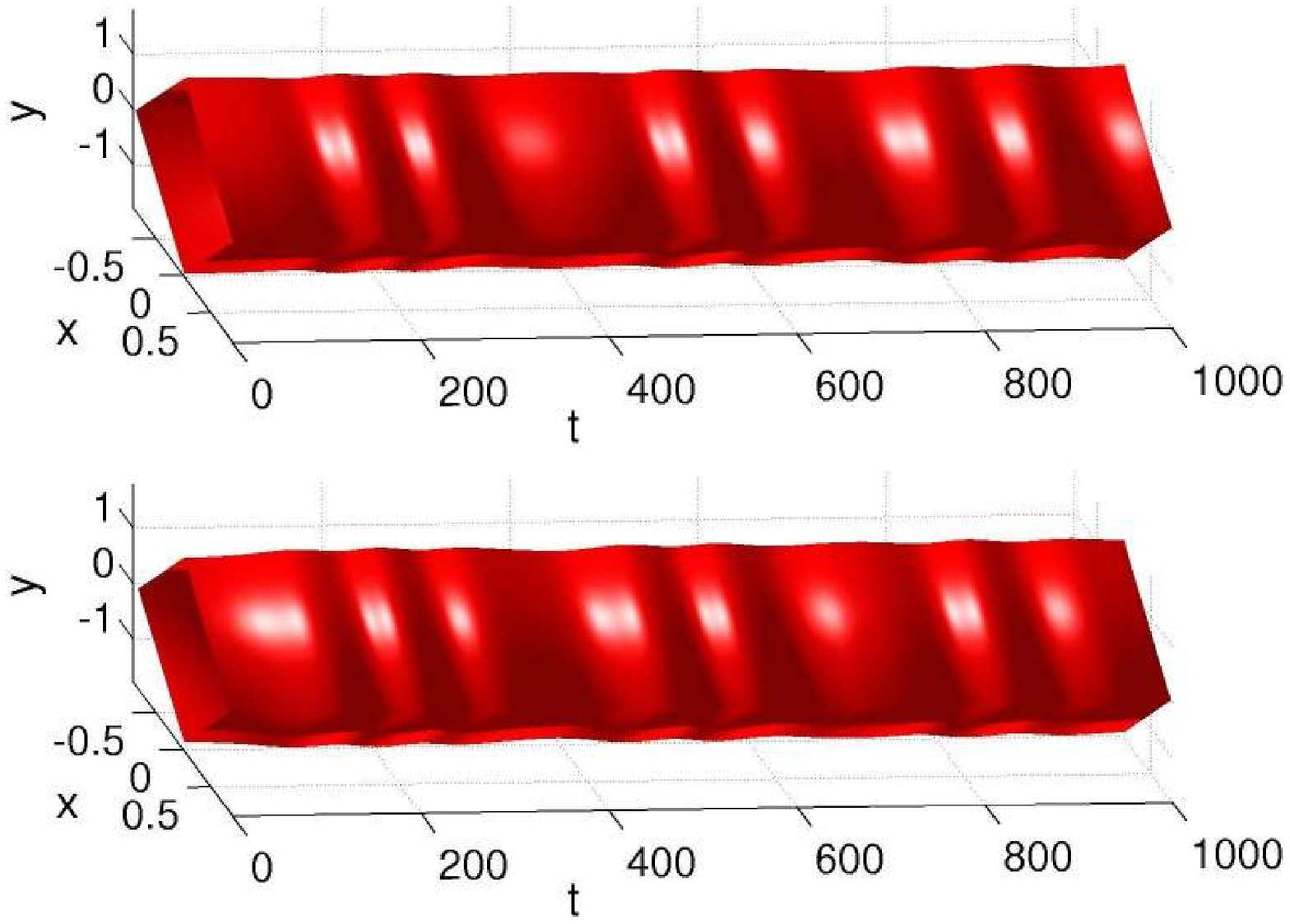}&
~\includegraphics[width=5.75cm,height=4.5cm,angle=0,clip]{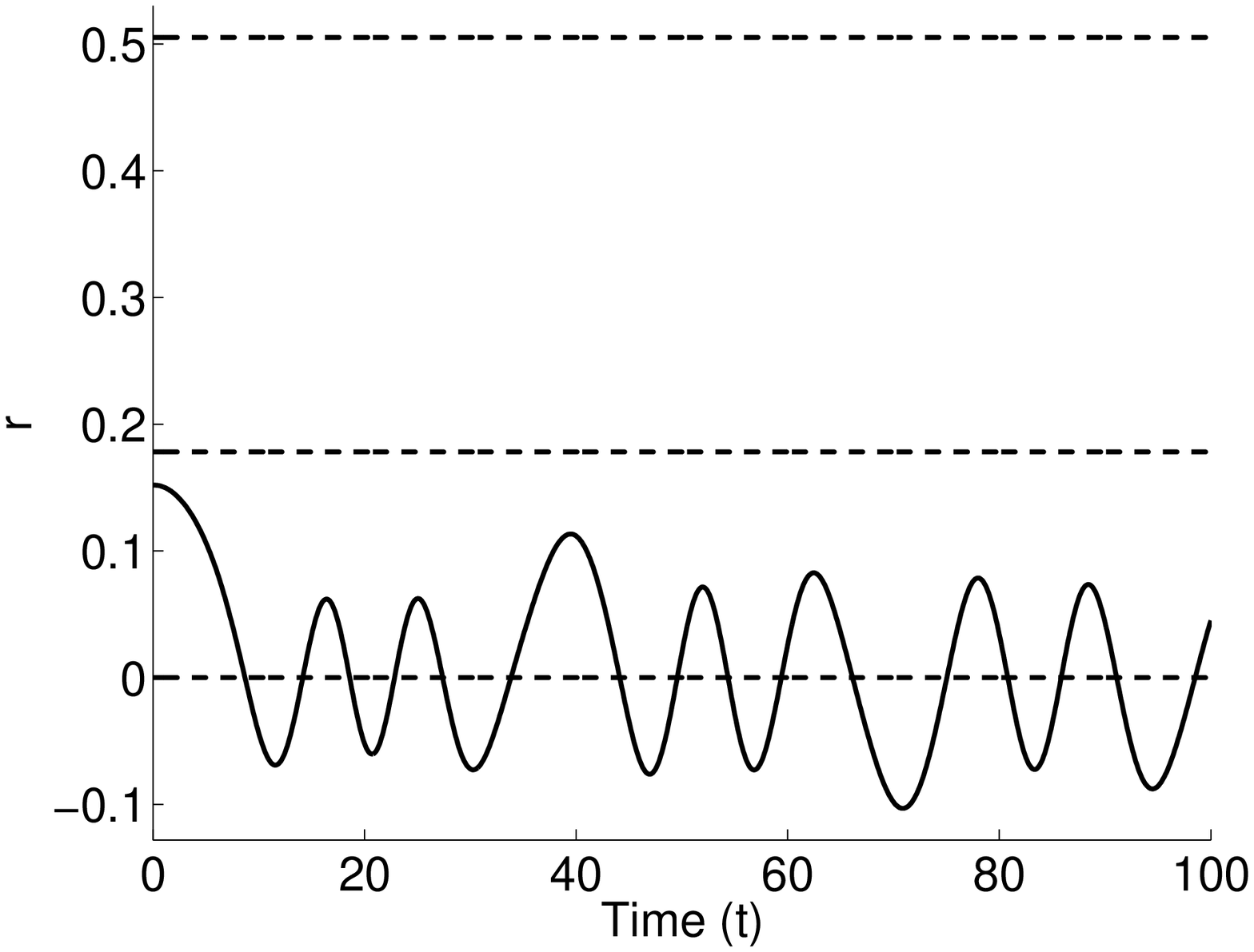}
\end{tabular}
\caption{Time-evolution plots for a perturbation of the unstable stationary
state from Fig.~\protect\ref{Solns2D1}. The top-row perturbation pushes
the solution towards the upper stable part of the curve, while the
bottom-row perturbation pushes it to the symmetric branch. The left
column shows the three-dimensional space-time evolution of
isodensity contours of the $U$ (respective top subpanels)
and $V$ (respective bottom subpanels) solutions,
while the right column shows  the
evolution of the asymmetry measure, $r$, from an unstable steady state
towards a stable one (both are denoted by dashed lines).}
\label{TimeEvo2D1}
\end{figure*}

\begin{figure*}[ht]
\begin{tabular}{cc}
 \includegraphics[width=5.75cm,height=4.5cm,angle=0,clip]{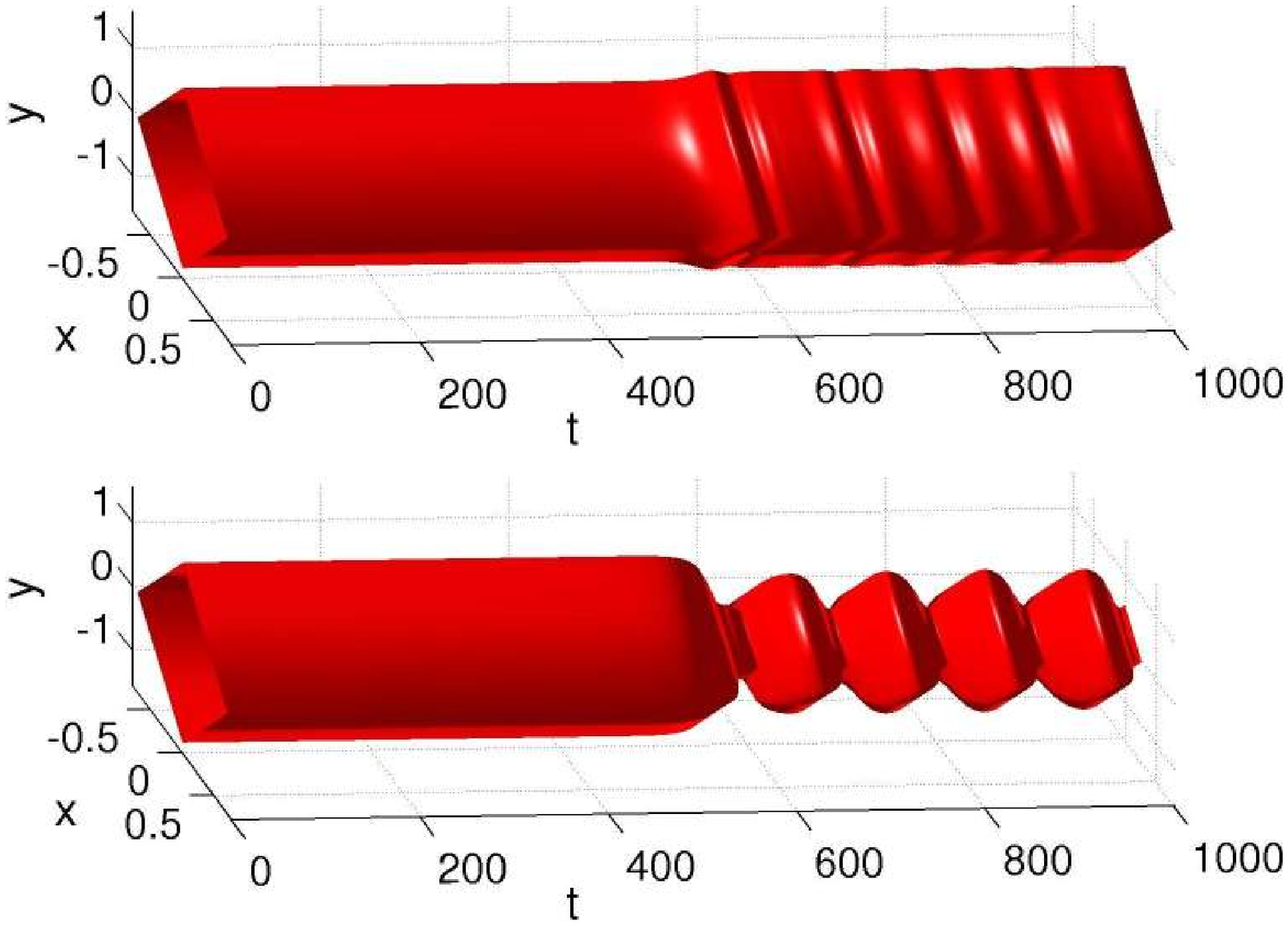}&
~\includegraphics[width=5.75cm,height=4.5cm,angle=0,clip]{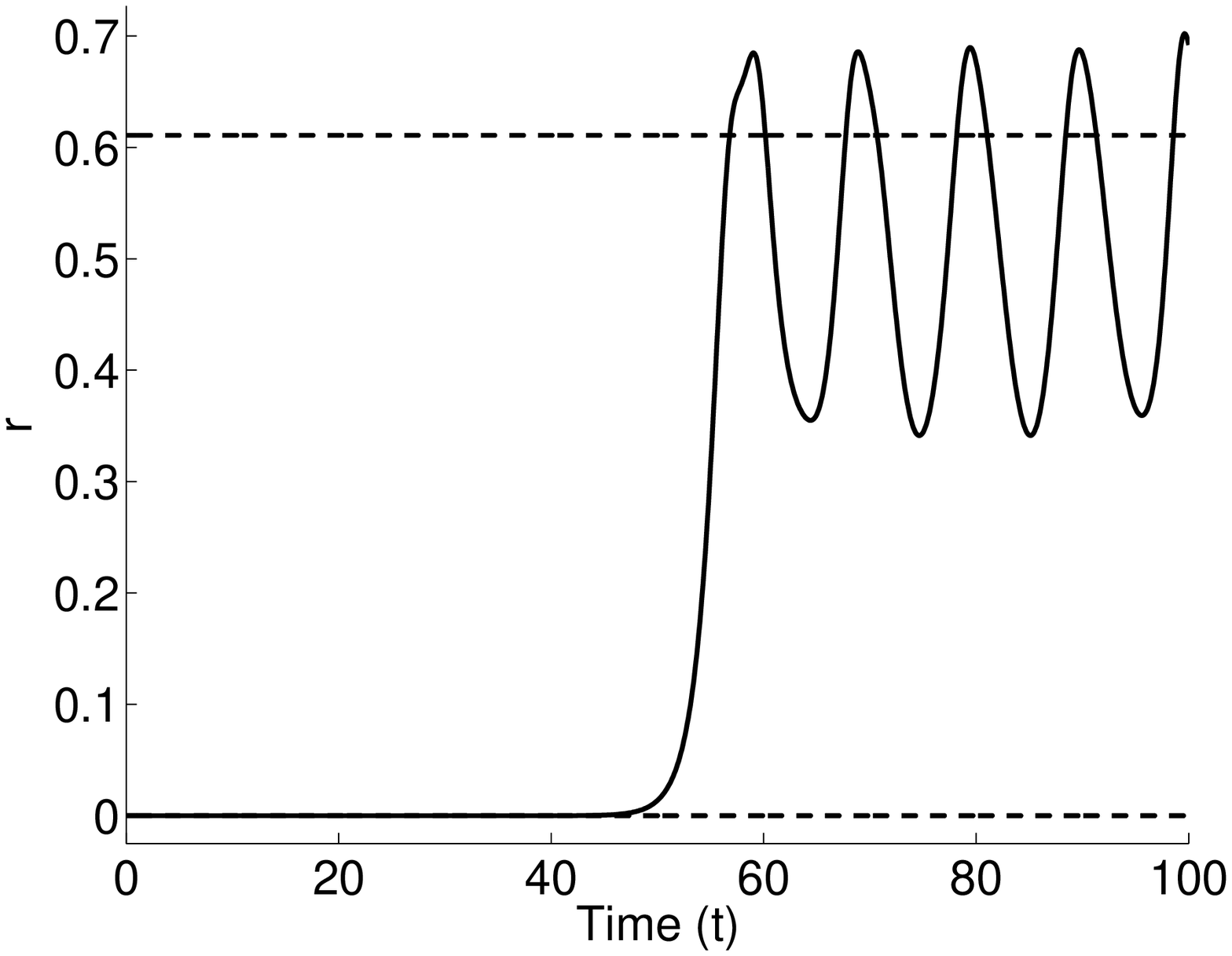}\\[2.0ex]
 \includegraphics[width=5.75cm,height=4.5cm,angle=0,clip]{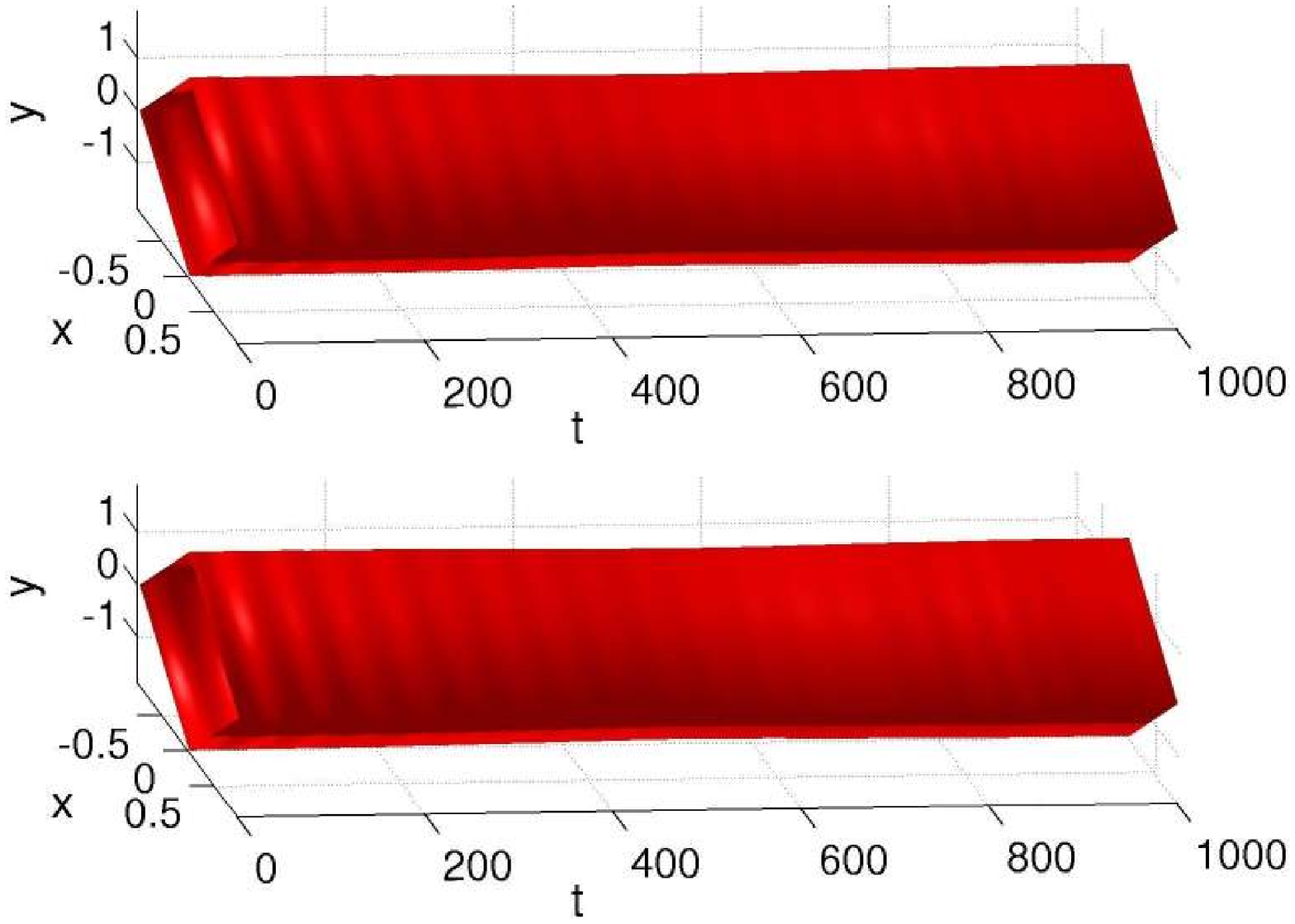}&
~\includegraphics[width=5.75cm,height=4.5cm,angle=0,clip]{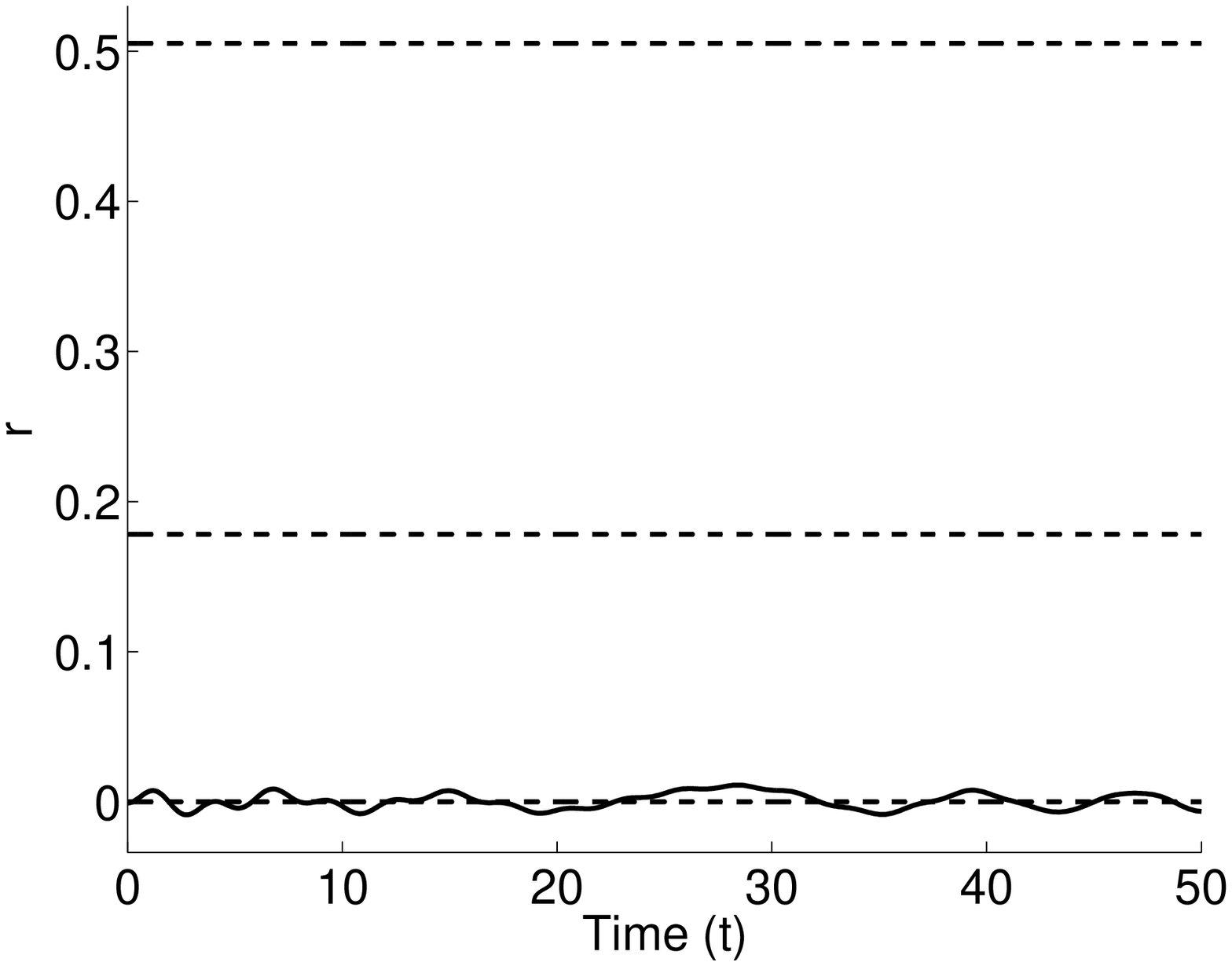}
\end{tabular}
\caption{Similar to the previous figure: the time-evolution plots for
perturbations of two unstable symmetric steady states. The top row is for a
solution with $E=1.5$, and the bottom row is for the solution from
Fig.~\protect\ref{Solns2D2}. In the top row, the perturbation pushes the solution
towards the upper asymmetric branch, while in bottom row the perturbation
initiates the evolution of the solution towards the stable symmetric
state. }
\label{TimeEvo2D2}
\end{figure*}

In the present work, motivated by the
earlier works \cite{akhmediev,chu,chu2}, which studied effects of
the linear coupling in various continuum optical models, and also the
analysis of the symmetry breaking of discrete two-component
solitons in the linearly coupled Ablowitz-Ladik system \cite{borisyang},
we consider a system of two DNLS equations coupled {\it solely} by linear
terms. In the optical setting, this would be the
above-mentioned discrete analog of a dual-core fiber
\cite{akhmediev}. Such a model has also been proposed as a means
for realization of all-optical switching in ultrashort
photonic-crystal couplers \cite{deangelis}. In BECs, it may model
the dynamics of two coupled hyperfine states (e.g., of $^{87}$Rb),
where we imply the use of Feshbach-resonance techniques \cite{feshbach}
to nullify the nonlinear interaction between the
components (by rendering the respective inter-species scattering
length equal to zero). At the same time, two spin states may be
linearly coupled due to a resonant spin-flipping radio-frequency
field, as mentioned above. In both media (optical and atomic), the
relevant model takes the following form:
\begin{equation}
\left\{
\begin{array}{c}
iU_{t}=K\epsilon \Delta _{2}U+KV+\left\vert U\right\vert ^{2}U \\[2.0ex]
iV_{t}=K\epsilon \Delta _{2}V+KU+\left\vert V\right\vert ^{2}V%
\end{array}%
\right. ,  \label{LinCoup}
\end{equation}
where $U=U(\vec{x},t)$ and $V=V(\vec{x},t)$ are the wave functions in BEC or
electric field envelopes in optics ($\vec{x}$ is realized as a set of
discrete coordinates), $\Delta _{2}$ is the discrete Laplacian, formed by
the centered difference in each of the relevant dimensions, $K$ is the
strength of the linear coupling between fields $U$ and $V$, and $\epsilon$
determines the couplings between adjacent sites of the lattice. Note that, for convenience,
the full coupling constant is defined as $K\epsilon$;
this convention will allow us to eliminate $K$ from the analysis presented below.

Our aim is to find and explore in detail the symmetry-breaking
bifurcation of the ground-state single-pulse solution of
Eqs.~(\ref{LinCoup}), similar to the earlier studies performed for
continuum models in Refs.~\cite{akhmediev,chu,chu2}. We will
address this problem analytically, by means of a variational
approximation (VA), and numerically, via computations of the
steady-state bifurcations and linear stability, as well as through
direct numerical simulations testing the stability or instability
of states under consideration. In this way, we obtain a
complete bifurcation diagram of the discrete model. A unique
advantage of performing this analysis in the discrete setting is
that the bifurcation diagram can be produced not only for
one-dimensional (1D), but also for two-dimensional (2D) lattices.
The latter is impossible in usual continuum models of the
cubic-NLS type because of the collapse instability \cite{Sulem}.
However, it was demonstrated in Ref.~\cite{pre}
(see also Ref.~\cite{dnls}) that the DNLS with sufficiently weak inter-site
coupling [i.e., small coefficient $\epsilon $ in Eqs.~(\ref{LinCoup})]
gives rise to \emph{stable} discrete 2D solitons. This permits us
to construct a bifurcation diagram for the 2D lattice and compare
it to the 1D counterpart. To the best of our knowledge, this is
the first time that a symmetry-breaking bifurcation is found and
analyzed in a 2D model with the cubic nonlinearity.

The paper is structured as follows. In Section 2, we present the model and
analytical results, based on the variational method. In Section 3, numerical
results are reported. Finally, in Section 4, we summarize our findings and
discuss directions for future work.

\section{The Model and Analytical Considerations}

In what follows, we seek steady-state (standing-wave) solutions to system (%
\ref{LinCoup}). We use the
standing-wave ansatz,
$$
\left\{
\begin{array}{rcl}
U\left(\vec{x},t\right) &=&\sqrt{K}\, u\left(\vec{x}\right)\,
\exp [-iK\left( \mu -2D\epsilon \right) t]  \\[2.0ex]
V\left(\vec{x},t\right) &=&\sqrt{K}\, v\left(\vec{x}\right)\,
\exp [-iK\left( \mu -2D\epsilon \right) t]
\end{array}
\right. ,
$$
where
$u(\vec{x})$ and $v(\vec{x})$ are real-valued functions,
$\mu$ is the chemical potential (propagation constant)
in the BEC (optics) model,
shifted by the constant $D$; the latter takes the values $D=1$ and $2$ for the 1D
and 2D cases, respectively.
Substituting these
expressions in Eqs.~(\ref{LinCoup}) leads to stationary equations, which, in the 1D
model, take the form:
\begin{equation}
\left\{
\begin{array}{c}
\mu u_{n}=\epsilon \overline{\Delta}_1 u_n +v_{n}+u_{n}^{3}\hfill \\[2.0ex]
\mu v_{n}=\epsilon \overline{\Delta}_1 v_n +u_{n}+v_{n}^{3}\hfill%
\end{array}%
\right. ,  \label{LinCoup1D}
\end{equation}
Where $\overline{\Delta}_1 w_n \equiv w_{n+1}+w_{n-1}$.
In the 2D case, the stationary equations are:
\begin{equation}
\left\{
\begin{array}{c}
\mu u_{n,m}=\epsilon \overline{\Delta}_2 u_{n,m} +v_{n,m}+u_{n,m}^{3} \\[2.0ex]
\mu v_{n,m}=\epsilon \overline{\Delta}_2 v_{n,m} +u_{n,m}+v_{n,m}^{3}%
\end{array}%
\right. ,  \label{LinCoup2D}
\end{equation}
Where $\overline{\Delta}_2 w_{n,m} \equiv w_{n+1,m}+w_{n-1,m}+w_{n,m+1}+w_{n,m-1}$.

Below, we aim to construct symmetric ($u=v$) and asymmetric
($u\neq v$) solutions of Eqs.~(\ref{LinCoup1D}) and
(\ref{LinCoup2D}). In order to develop an analytical approximation
to the solutions, we resort to the variational method
\cite{Progress}. To this end, we notice that Eqs.~(\ref{LinCoup1D})
and (\ref{LinCoup2D}) can be derived from the
following Lagrangians:
\begin{eqnarray}
\displaystyle
L_{\mathrm{1D}}&=&
\sum\limits_{n=-\infty }^{\infty }
\left[ {-{\frac{{\mu }}{2}}}\left( {u_{n}^{2}+v_{n}^{2}}\right)
+{\frac{1}{4}}\left( {u_{n}^{4}+v_{n}^{4}}\right)\right. \nonumber\\[1.0ex]
&&
\left.
\phantom{\frac{1}{1}}
+u_{n}v_{n}
\epsilon \left( {u_{n+1}u_{n}+v_{n+1}v_{n}}\right) \right] ,
\label{Lagrangian1D}
\end{eqnarray}
\begin{eqnarray}
\displaystyle
&&L_{\mathrm{2D}}=\sum\limits_{m,n=-\infty }^{\infty }\left[ {-\frac{{\mu }%
}{2}}\left( {u_{n,m}^{2}+v_{n,m}^{2}}\right) +
\right. \nonumber\\[1.0ex]
\displaystyle
&&
{\frac{1}{4}}\left( {u_{n,m}^{4}+v_{n,m}^{4}}\right)
+u_{n,m}^{4}v_{n,m}^{4}
+\epsilon (u_{n+1,m}u_{n,m}+ \nonumber\\[1.0ex]
&&
\left.
\phantom{\frac{1}{1}}
u_{n,m+1}u_{n,m}+v_{n+1,m}v_{n,m}+v_{n,m+1}v_{n,m})
\right] .
 \label{Lagrangian2D}
\end{eqnarray}
Further, we employ natural \textit{ans\"{a}tze}, $\{u_{n},v_{n}\}=\{A,B%
\}e^{-\lambda |n|}$ and $\{u_{n,m},v_{n,m}\}=\{A,B\}e^{-\lambda
|n|}e^{-\lambda |m|}$,
with free constants $A$, $B$ and $\lambda >0$, in the
1D and 2D cases, respectively. This choice is motivated both by the
exponential decay of the solutions' tails far from the soliton's center and
by the fact that
only this type of the trial functions makes the approximation 
straightforwardly tractable \cite{MIW}.
Plugging the \textit{ans\"{a}%
tze} into Eqs.~(\ref{Lagrangian1D}) and (\ref{Lagrangian2D}) and
analytically evaluating the resulting geometric series, we arrive at the
following expressions for the effective Lagrangians:
\begin{eqnarray}
&&L_{\mathrm{1D}}=\left[ AB-{\frac{{\mu }}{2}}\left( {A^{2}+B^{2}}\right) %
\right] \coth \lambda
\nonumber\\[1.0ex]
&&
~~~~
+{\frac{1}{4}}\left( {A^{4}+B^{4}}\right) \coth \left(2\lambda \right)
+\epsilon \left( {A^{2}+B^{2}}\right) \mathrm{cosech}\lambda~~~~~~~~
\\[2.0ex]
&&L_{\mathrm{2D}}=\left[ AB-{\frac{{\mu }}{2}}\left( {A^{2}+B^{2}}\right) %
\right] \coth ^{2}\lambda
\nonumber\\[1.0ex]
&&
~~~~~~
+{\frac{1}{4}}\left( {A^{4}+B^{4}}\right) \coth^{2}2\lambda
\nonumber\\[1.0ex]
&&
~~~~~~
+2\epsilon \left( {A^{2}+B^{2}}\right) \left( \mathrm{cosech}%
\lambda \right) \coth \lambda .
\end{eqnarray}
The static form of the
Euler-Lagrange equations following from here, $\partial L_{\mathrm{1D,2D}%
}/\partial \left( \lambda ,A,B\right) =0$, is
\begin{eqnarray}
{\frac{\mu }{2}}\left( {A^{2}+B^{2}}\right) \mathrm{cosech}^{2}\lambda -{%
\frac{1}{2}}\left( {A^{4}+B^{4}}\right) \mathrm{cosech}^{2}2\lambda
\nonumber\\[1.0ex]
-AB\mathrm{cosech}^{2}\lambda
\epsilon \left( {A^{2}+B^{2}}\right) \mathrm{cosech}\lambda \coth \lambda
=0,
\nonumber\\[2.0ex]
-\mu A\coth \lambda +A^{3}\coth 2\lambda +B\coth \lambda +2\epsilon A\mathrm{%
cosech}\lambda  &=&0,
\nonumber\\[2.0ex]
-\mu B\coth \lambda +B^{3}\coth 2\lambda +A\coth \lambda +2\epsilon B\mathrm{%
cosech}\lambda  &=&0,
\nonumber
\end{eqnarray}
for the 1D case, and
\begin{eqnarray}
&~&\mu \left( {A^{2}+B^{2}}\right) \coth \lambda \mathrm{cosech}^{2}\lambda
\nonumber\\[1.0ex]
&-&\left( {A^{4}+B^{4}}\right) \coth 2\lambda \mathrm{cosech}^{2}2\lambda
-2AB\coth \lambda \mathrm{cosech}^{2}\lambda
\nonumber\\[1.0ex]
&-&2\epsilon \left( { A^{2}+B^{2}}\right) \left( {\mathrm{cosech}\lambda \coth ^{2}\lambda +%
\mathrm{cosech}^{3}\lambda }\right) =0,
\nonumber\\[2.0ex]
&-&\mu A\coth ^{2}\lambda +A^{3}\coth ^{2}2\lambda +B\coth ^{2}\lambda
+\nonumber\\[1.0ex]
&&4\epsilon A\mathrm{cosech}\lambda \coth \lambda  =0,
\nonumber\\[2.0ex]
&-&\mu B\coth ^{2}\lambda +B^{3}\coth ^{2}2\lambda +A\coth ^{2}\lambda
\nonumber\\[1.0ex]
&&+4\epsilon B\mathrm{cosech}\lambda \coth \lambda  =0,
\nonumber
\end{eqnarray}
for the 2D case.
These equations were solved for $A$, $B$, and $\lambda $ and will
be compared, in the next section, to the full numerical solution of
Eqs.~(\ref{LinCoup1D}) and (\ref{LinCoup2D}).

Another analytically tractable case corresponds to the anti-continuum limit
of $\epsilon =0$. For the symmetric branch, we then have $u_{n}=v_{n}=0$ or $%
u_{n}=v_{n}=\sqrt{\mu -1}$, while for the asymmetric branch, one needs to
solve a system of algebraic equations, $\mu u_{n}=v_{n}+u_{n}^{3}$,$~\mu
+1=u_{n}^{2}+u_{n}v_{n}+v_{n}^{2}$. The solution is shown in Fig.~\ref{Eps0},
which displays the respective symmetry-breaking bifurcation by
means of a plot of the asymmetry measure, $r\equiv
(E_{1}-E_{2})/(E_{1}+E_{2})$, versus half the total norm,
$E=(E_{1}+E_{2})/2$, where $\left\{ E_{1},E_{2}\right\}
=\sum_{n}\left\{ u_{n}^{2},v_{n}^{2}\right\} $ are the norms of
the two components of the solution (in the BEC model, they are
proportional to the number of atoms in the two atomic states,
while in the optical setting they measure the total power of the
beams in the two coupled lattices). As seen from the figure, the
pitchfork bifurcation is supercritical in this limit (this will
be compared to typical behavior for finite $\epsilon $ below).

\section{Numerical Methods and Results}

Numerical solutions to Eqs.~(\ref{LinCoup1D}) and (\ref{LinCoup2D}) were
found by using
the method of the pseudo-arclength continuation \cite%
{arclength1,arclength2}. Another typical approach to solving systems of
nonlinear equations for various values of a control parameter relies upon
parameter continuation; however, this method fails for solution-parameter
pairings where the resulting Jacobian is singular. On the other hand, the
pseudo-arclength continuation addresses this problem by introducing an extra
pseudo-arclength parameter, and including an additional equation into the
system, which makes the solution and control parameter dependent upon the
pseudo-arclength. Calling the pseudo-arclength parameter $s$, the additional equation,
$F(u,v,\mu ,s)=0$%
, must be chosen such that $F({\bar{u}},{\bar{v}},{\bar{\mu}},s=0)=0$, where
$({\bar{u}},{\bar{v}},{\bar{\mu}})$ is a solution of Eqs.~(\ref{LinCoup1D})
or (\ref{LinCoup2D}).
Thus, for $F$ we used $F(u,v,\mu
,s)=\left\vert {u-\bar{u}}\right\vert ^{2}+\left\vert {v-\bar{v}}\right\vert
^{2}+\left\vert {\mu -\bar{\mu}}\right\vert ^{2}-s^{2}$.

Once the steady states were identified, their stability was examined in the
framework of linear stability analysis by
substituting a perturbed solution, in the
form of
\begin{equation}
\left\{
\begin{array}{rcl}
U(\vec{x},t)&=&e^{-i\mu t}\left[ u(\vec{x})+
a(\vec{x})\,e^{\lambda t}+b^{\ast}(\vec{x})\,e^{\lambda ^{\star }t}\right] ,
\\[2.0ex]
V(\vec{x},t)&=&e^{-i\mu t}\left[ v(\vec{x})+
c(\vec{x})\,e^{\lambda t}+d^{\ast}(\vec{x})\,e^{\lambda ^{\star }t}\right] ,
\end{array}
\right.
\label{PertAnsatz}
\end{equation}
in Eqs.~(\ref{LinCoup}), the asterisk standing for complex conjugation.
The linearized equations for the perturbation eigenmodes $a, b, c, d$ were solved
numerically, yielding eigenvalues $\lambda $ associated with them.

The results are summarized in Figs.~\ref{Branches} and \ref{Branches2D}, for
the 1D and 2D cases, respectively. In the 1D case, the symmetric branch,
with $r=0$, is stable for $E<3.82$. Beyond this critical point, it becomes
unstable through a \textit{subcritical} pitchfork bifurcation, due to its
collision with two unstable asymmetric branches. The VA predicts this
critical point at $E\approx 3.92$, in good agreement with the numerical
findings. Further, at another critical value, $E=3.166$ (the corresponding
VA prediction is $E\approx 3.128$) the unstable asymmetric branches turn
back as stable ones, through a saddle-node bifurcation. Between the two
critical points, both the symmetric branch and the outer asymmetric one are
stable, hence there exists 
a region of bistability. To additionally demonstrate the
accuracy of the VA, Fig.~\ref{Solns} presents comparison of the solution
profiles at $E=3.4$, together with the spectral plane, $(\mathrm{Re}(\lambda
),\mathrm{Im}(\lambda ))$, for the corresponding eigenvalues, $\lambda
\equiv \mathrm{Re}(\lambda )+i\mathrm{Im}(\lambda )$. Recall that in
Hamiltonian systems, such as the one considered here, if $\lambda $ is an
eigenvalue, so are also $-\lambda $, $\lambda ^{\star }$ and $-\lambda
^{\star }$, hence, if an eigenvalue with a nonzero real part exists, then
the system will be linearly unstable.

For those 1D solutions that are linearly unstable due to a real
eigenvalue, such as the unstable asymmetric solutions for
$3.166<E<3.82$ and the symmetric ones for $E>3.82$, we have
examined their evolution in Fig.~\ref{TimeEvo} by means of direct
simulations of  Eq.~(\ref{LinCoup}). In the case of
the unstable asymmetric branch, the result of the evolution
depends on the nature of the corresponding perturbation, due to
the presence of the bistability in the corresponding parameter
range: the solution ends up oscillating around either the stable
asymmetric solution, or the stable symmetric one. The evolution of
the unstable symmetric branch naturally results in oscillations
around the stable asymmetric profile, which represents the ground
state in that case.

An important observation in comparing Figs.~\ref{Eps0} and \ref{Branches} is
that the bifurcation found in the anti-continuum limit shown in
Fig.~\ref{Eps0}\ is definitely supercritical, unlike the weakly subcritical one
in Fig.~\ref{Branches}. This indicates that the character of the bifurcation
changes from subcritical to supercritical pitchfork with the increase of
discreteness, i.e.,\ decrease of $\epsilon $. This transition should
eliminate the unstable asymmetric branches. In accordance with this
expectation, we have found that the unstable asymmetric solutions exist only
for $\epsilon >0.35$, in the 1D case.

We now turn to the results for the 2D model collected in
Fig.~\ref{Branches2D}.
Here, the results are even more interesting, for a number of
reasons. On the one hand, there is no continuum analog to the bifurcation
diagram, as the respective 2D continuum solutions are always unstable to
collapse. Discreteness is well-known to arrest collapse \cite{pre,arrest},
generating branches of potentially stable localized solutions for
sufficiently small values of the inter-site coupling constant (at a given
chemical potential), or for sufficiently large chemical potential (at a
given inter-site coupling), in one-component models \cite{pre}. Furthermore,
in the 2D case, for a given value of the norm, there are two coexisting
symmetric solutions, one (taller and narrower) with a larger chemical
potential, which is stable, and one (shorter and wider) with a smaller
chemical potential, which is unstable. As Fig.~\ref{Branches2D} implies, the
symmetry-breaking weakly subcritical pitchfork bifurcation \textit{typically}
occurs from the stable branch of the symmetric solution, the corresponding
critical point in Fig.~\ref{Branches2D} being $E\approx 1.45$. 
Similarly to
the 1D case, 
there is also a saddle-node bifurcation between the unstable
and stable asymmetric branches, which occurs at $E\approx 1.411$ and is
responsible for the turning point.

In the 2D model too, the VA accurately captures the trends of the numerical
results, even though the less accurate nature of the 2D ansatz prevents a
quantitative matching of the resulting bifurcation diagrams. Detailed
profiles of the numerical solutions and their VA-predicted counterparts are
shown in Figs.~\ref{Solns2D1} and \ref{Solns2D2}, for the asymmetric and
symmetric branches respectively, at $E=1.435$. Two additional remarks are in
order here. Firstly, similar to the 1D case, the 2D bifurcation changes
character from weakly subcritical (as observed in Fig.~\ref{Branches2D}) to
supercritical (as seen in the anti-continuum limit of $\epsilon =0$) at $%
\epsilon \approx 0.19$. On the other hand, as $\epsilon $ further increases,
the stable part of the symmetric branch (in Fig.~\ref{Branches2D}) shrinks
and eventually disappears at $\epsilon >0.29$.

Finally, we have again examined the dynamics of linearly unstable solutions,
upon appropriate perturbations, through direct simulations,
see Figs.~\ref{TimeEvo2D1} and \ref{TimeEvo2D2}.
In the former figure, we have explored
how the bistability, which is shown for a range of $E$
in Fig.~\ref{Branches2D},
``kicks" the unstable asymmetric solution either in
the direction of its stable asymmetric counterpart, or towards the
stable symmetric solution [it is relevant to stress here, in
connection with Fig.~\ref{Branches2D}, that, while the symmetric
solution has a norm threshold at $E\approx 1.425$, the stable
asymmetric solution can, in principle, be found at \textit{lower}
norms than the symmetric one, thus allowing the system to
effectively decrease its ``excitation threshold" \cite{wein}]. In
the latter figure, we consider the unstable symmetric branch, both
when a stable symmetric branch does not exist, in which case the
solution becomes asymmetric (top row), and when a stable symmetric
branch does exist, in which case the system evolves towards that
solution (bottom row).



\section{Conclusions}

In this work, we have introduced the model based on two linearly coupled
lattices with the cubic nonlinearity, and investigated its dynamical
properties in detail. We have demonstrated that, in a number of respects,
the discrete system emulates its continuum 1D counterpart, analyzed earlier
in Refs.~\cite{akhmediev,chu,chu2}. On the other hand, the lattice model
gives rise to novel features. Even in the 1D setting, varying the strength
of the lattice inter-site coupling may be used to switch the character of
the bifurcation from subcritical to supercritical pitchfork, as the coupling
gets weaker, and the anti-continuum limit is approached. In the more
interesting 2D setting, our work is the first manifestation, to the best of
our knowledge, of the existence of such a bifurcation diagram, since in the
continuum limit the symmetry-breaking models with the self-focusing
nonlinearity are irrelevant due to the collapse instability. Furthermore,
the discreteness induces the presence of both stable and unstable branches
of symmetric solutions, thus enriching the bifurcation diagram. In the 2D
case, not only is it possible for weaker lattice coupling to turn the
bifurcation from subcritical to supercritical, but it is also possible for
the lattice (when the bifurcation is subcritical) to possess a lower
excitation threshold for asymmetric states than for the symmetric ones. All
of these features demonstrate critical modifications that the discreteness
imposes on the well-known symmetry-breaking picture in continuum models.

It might be quite interesting to examine similar features in 3D and compare
the results with their 2D and 1D counterparts. Of perhaps even more physical
interest, especially in terms of coupled hyperfine states in BECs, would be
to add nonlinear coupling between the lattice components, of the
cross-phase-modulation type, to the linear coupling considered here. It
would be particularly interesting to examine how the symmetry-breaking
phenomenology is affected by the gradual increase of such a coupling. This
study is currently in progress and the results will be reported elsewhere.

\end{document}